\documentclass[12pt,epsf]{article}
\usepackage{verbatim}
\usepackage{anyfontsize}
\usepackage{dsfont}
\usepackage{tikz}
\usepackage{sidecap}
\sidecaptionvpos{figure}{t}
\usepackage{subfigure}
\usepackage{enumitem}
\usepackage[english]{babel}
\usepackage{enumitem}  

\usepackage{amssymb,amsmath}
\usepackage{graphicx, xcolor, varwidth}
\usepackage{setspace}
\usepackage[permil]{overpic}
\usepackage{cite}
\usepackage{mathtools}

\DeclareSymbolFont{matha}{OML}{txmi}{m}{it}% txfonts
\DeclareMathSymbol{v}{\mathord}{matha}{118}

\usepackage[framemethod=default]{mdframed}
\newmdenv[skipabove=7pt,
skipbelow=7pt,
rightline=false,
leftline=false,
topline=false,
bottomline=false,
backgroundcolor=blue!10,
linecolor=blue,
innerleftmargin=5pt,
innerrightmargin=5pt,
innertopmargin=5pt,
innerbottommargin=5pt,
leftmargin=0cm,
rightmargin=0cm,
linewidth=4pt]{bBox}

\colorlet{darkblue}{blue!70!black}
\colorlet{darkgreen}{green!70!black}

\usepackage[colorlinks=true,urlcolor=darkblue,linktocpage=true,linkcolor=darkblue,citecolor=darkblue]{hyperref}
\hypersetup{ 
colorlinks=true, 
linkcolor=blue, 
citecolor=magenta, 
}

\numberwithin{equation}{section}
\DeclareMathSymbol{v}{\mathord}{matha}{118}
\newcommand{\be}{\begin{equation}}
\newcommand{\ee}{\end{equation}}
\newcommand{\bea}{\begin{eqnarray}}
\newcommand{\eea}{\end{eqnarray}}
\newcommand{\bear}{\begin{eqnarray}}
\newcommand{\eear}{\end{eqnarray}}
\newcommand{\beas}{\begin{eqnarray*}}
\newcommand{\p}{\partial}
\newcommand{\eeas}{\end{eqnarray*}}
\newcommand{\ba}{\begin{array}}
\newcommand{\ea}{\end{array}}

\def\ba#1\ea{\begin{align}#1\end{align}}
\def\bs#1\es{\begin{split}#1\end{split}}

\renewcommand{\r}{\rho}
\newcommand{\br}{\bar{\rho}}

\newcommand{\tr}{\operatorname{tr}}
\newcommand{\pd}[2][1]{\ifnum#1=1 \frac{\partial}{\partial {#2}} \else
  \frac{\partial^#1}{\partial {#2}^{#1}}\fi}
\newcommand{\dpd}[2][1]{\ifnum#1=1 \dfrac{\partial}{\partial {#2}} \else
  \frac{\partial^#1}{\partial {#2}^{#1}}\fi}
\newcommand{\td}[2][1]{\ifnum#1=1 \frac{d}{d{#2}} \else
  \frac{d^#1}{d{#2}^{#1}}\fi}

\renewcommand{\c}{\tilde{c}}
\renewcommand{\(}{\left(}
\renewcommand{\)}{\right)}
\renewcommand{\[}{\left[}
\renewcommand{\]}{\right]}

\newcommand{\nbox}{{\,\lower0.9pt\vbox{\hrule \hbox{\vrule height 0.2 cm \hskip 0.19 cm \vrule height 0.2 cm}\hrule}\,}}
\newcommand{\Tr}{\ {\rm Tr}\ }

\newcommand{\eg}{{\it e.g.,}\ }

\def\O{{\cal O}}

\newcommand{\bA}{\mathcal{A}}

\newcommand{\N}{{\cal N}}
\newcommand{\CFTUV}{\text{CFT}_{\text{UV}}}
\newcommand{\CFTIR}{\text{CFT}_{\text{IR}}}



\begin{document}
\begin{spacing}{1.3}
\begin{titlepage}
  % end of \vbox

\begin{center}
{\Large 
\vspace*{6mm}

RG Flows with Global Symmetry Breaking\\ 
and \\
 Bounds from Chaos

}

\vspace*{6mm}

Sandipan Kundu

\vspace*{6mm}

\textit{Department of Physics and Astronomy,
\\ Johns Hopkins University,
Baltimore, Maryland, USA\\}

\vspace{6mm}

{\tt \small kundu@jhu.edu}

\vspace*{6mm}
\end{center}

\begin{abstract}

We discuss general aspects of renormalization group (RG) flows between two conformal fixed points in 4d with a broken continuous global symmetry in the UV. Every such RG flow can be described in terms of  the dynamics of Nambu-Goldstone bosons of broken conformal and global symmetries. We derive the low-energy effective action that describes this class of RG flows  from basic symmetry principles. We view the theory of Nambu-Goldstone bosons as a theory in anti-de Sitter space with the flat space limit. This enables an equivalent CFT$_3$ formulation of these 4d RG flows in terms of spectral deformations of a generalized free CFT$_3$. We utilize this dual description to impose further constraints on the low energy effective action associated with unitary  RG flows in 4d by invoking the chaos bound in 3d. This approach naturally provides a set of independent monotonically decreasing $C$-functions for 4d RG flows with global symmetry breaking by explicitly relating 4d $C$-functions with   certain out-of-time-order correlators that diagnose chaos in 3d. We also comment on a more general connection between RG and chaos in QFT.

\end{abstract}

\end{titlepage}
\end{spacing}

\vskip 1cm
\setcounter{tocdepth}{2}  
\tableofcontents

\begin{spacing}{1.3}

%%%%%%%%%%%%%%%%%%%%%%%%%%%%%%%%%%%%%%%%%
\section{Introduction}
The renormalization group (RG) and quantum chaos are two fundamentally important but distinct phenomena in quantum field theory (QFT) with some similar qualitative features. For example, both RG flows and semiclassical chaos exhibit certain universal positivity and monotonicity properties in generic quantum systems. Over the years, a great deal of progress has been made on understanding such general features of both RG and chaos, however, any connection between their positivity and monotonicity properties has never been established. This is not surprising since the underlying physics associated with RG and chaos are believed to be different. Nevertheless, in this paper we present a precise but indirect connection between RG and semiclassical chaos by considering a rather general class of RG flows in 4d. This also provides a tool to constrain unitary RG flows by utilizing the {\it chaos bound} of Maldacena, Shenker, and Stanford \cite{Maldacena:2015waa}.

Most physical systems, when viewed at different energy scales, admit descriptions in terms of completely different degrees of freedom. The  RG is a concrete realization of this phenomenon in QFT. It is a systematic coarse-graining procedure that identifies relevant long-distance degrees of freedom of a given quantum theory. Conformal field theories (CFTs) play a central role in RG since it is long believed that fixed points of RG flows are CFTs.\footnote{In $d>4$, a CFT can flow to a fixed point which is scale-invariant but non-conformal \cite{Cordova:2015fha}. However, in this paper, we will only consider 4d RG flows between two CFTs.}  

On physical grounds, it is expected that all RG flows should  be irreversible.  Consider a $\CFTUV$ which is deformed by adding a relevant (or marginally relevant) operator that breaks conformal symmetry.\footnote{There are RG flows in which conformal symmetry is broken spontaneously. The same discussion applies for such RG flows as well.} This triggers an RG flow that ends at $\CFTIR$. The irreversibility requires that any RG flow that starts from $\CFTIR$ and ends at $\CFTUV$ must be forbidden. A closely related but not exactly equivalent statement is that there exist real positive definite {\it $C$-functions} on the space of couplings with the following properties:  (i) $C$ decreases monotonically under RG flows, (ii) at the fixed points of the RG flow, $C$ is constant and independent of the energy scale.  Moreover, the value of a $C$-function at fixed points depends only on $\CFTUV$ and $\CFTIR$, respectively.  The existence of a $C$-function  necessarily implies irreversibility of RG flows when it  interpolates between some central charge of  $\CFTUV$ and $\CFTIR$. Such a $C$-function was first found by Zamolodchikov in 1986 for any unitary, Lorentz invariant QFT in 2d establishing the irreversibility of 2d RG flows \cite{Zamolodchikov:1986gt}. In 4d, a $C$-function was found by Komargodski and Schwimmer in 2011 that interpolates between the Euler central charges in the ultraviolet and the infrared \cite{Komargodski:2011vj} (see also \cite{Komargodski:2011xv}). This proved Cardy's conjecture \cite{Cardy:1988cwa} $\Delta a= a_{\rm UV}-a_{\rm IR}> 0$ establishing that all unitary RG flows are irreversible in 4d.\footnote{A general proof of the RG irreversibility is still missing in 6d (for attempts see \cite{Elvang:2012st,Kundu:2019zsl}). On the other hand, the 6d $a$-theorem has been established for all 6d flows that preserve $(2, 0)$ supersymmetry in \cite{Cordova:2015vwa}. The proof was later extended to RG flows of $(1, 0)$ SCFTs onto the tensor branch in \cite{Cordova:2015fha}. However, a proof of the $a$-theorem for RG flows of $(1, 0)$ SCFTs onto the Higgs branch is still an open problem even though there is strong evidence in favor it \cite{Heckman:2015axa,Heckman:2016ssk, Heckman:2018jxk}.  } 

\begin{figure}
\centering
\includegraphics[scale=0.37]{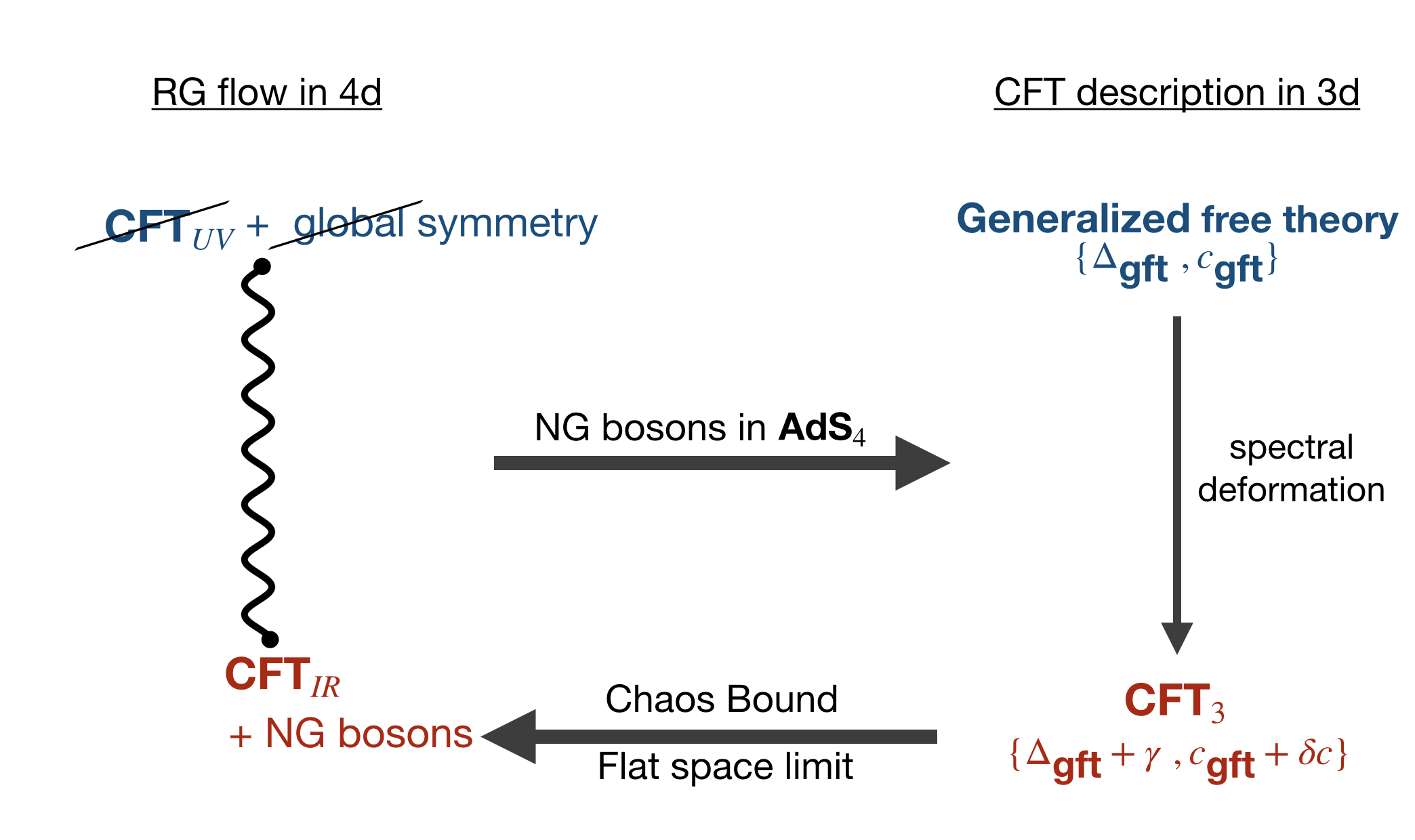}
\caption{ \label{intro} \small Every RG flow with global symmetry breaking can be described by the effective action of Nambu-Goldstone (NG) bosons of broken conformal and global symmetries. Any such RG flow in 4d has a dual CFT$_3$ description  where the dual CFT$_{3}$ is obtained by deforming operator dimensions and OPE coefficients of  a 3d  generalized free theory. The 3d chaos bound then imposes constraint on the effective action of the NG bosons. }
\end{figure}

In many 4d RG flows (\eg supersymmetric RG flows), the breaking of conformal symmetry is accompanied by the breaking of some other global symmetries of $\CFTUV$. In this paper, we consider RG flows between two conformal fixed points in 4d in which conformal symmetry and a continuous global symmetry are broken in the UV.  Our main argument can be briefly summarized as follows:
\begin{enumerate}
\item {Many general features of these RG flows, such as irreversibility and positivity, can be studied by analyzing the effective action of Nambu-Goldstone (NG) bosons of  broken conformal and global symmetries. By extending the argument of  \cite{Komargodski:2011vj}, we show that the general form of the effective action that describes 4d RG flows with global symmetry breaking  is completely fixed from symmetries. The effective action makes it obvious that the proof of the $a$-theorem remains unaffected even when global symmetries are broken. }
\item{Next, by following the framework of \cite{Kundu:2019zsl} we analyze the flat space effective theory of NG bosons by viewing it as a theory in anti-de Sitter (AdS) space with finite but large radius $R_{\rm AdS}$ and then take  the flat space limit $R_{\rm AdS}\rightarrow \infty$. This provides an alternative description of this class of 4d RG flows in terms of {\it spectral deformations of a generalized free CFT} in 3d.\footnote{Alternatively, one can combine the first two steps by imagining the RG flow between $\CFTUV$ and $\CFTIR$ is taking place in AdS$_4$ with $R_{\rm AdS}\rightarrow \infty$. These two interpretations are completely equivalent in the leading order of the effective action (up to four-derivative interactions). However, in general two interpretations may differ at higher derivative order.}}
\item{Finally, we utilize this dual description to derive positivity conditions for the effective action by invoking the chaos bound \cite{Maldacena:2015waa,Afkhami-Jeddi:2016ntf,Kundu:2020gkz} in the dual CFT$_3$. In particular, the chaos bound in 3d implies the $a$-theorem in 4d.\footnote{This connection was already noticed in \cite{Kundu:2019zsl}.} Furthermore, the 3d chaos bound provides a natural basis for constructing a set of 4d $C$-functions for RG flows with  global symmetry breaking.}
\end{enumerate}
Our approach, as summarized in figure \ref{intro},  connects RG and quantum chaos, albeit in different spacetime dimensions. 

\subsection{RG Flows with Global Symmetry Breaking}
In \cite{Komargodski:2011vj} Komargodski and Schwimmer taught us how every RG flow can be described in terms of a spontaneous breaking of conformal symmetry. We consider a more general class of RG flows in 4d where $\CFTUV$ has some global symmetry $G$, where $G$ is a compact Lie group.  The conformal symmetry and the global symmetry of $\CFTUV$ are broken either spontaneously or explicitly. This triggers an RG flow that preserves some subgroup $H$ of $G$. Following \cite{Komargodski:2011vj}, we argue that every such RG flow can be described as spontaneous breaking of conformal and global symmetries. The spontaneously broken conformal symmetry generates a {\it massless} NG boson -- the {\it dilaton} $\phi$.  The dilaton is accompanied by $N=\mbox{dim}\ G/H$ additional massless NG bosons $\xi_i$  arising from the spontaneous breaking of  the global symmetry.   So, in general the low energy theory consists of $\CFTIR$ and $(N+1)$ massless scalars $\phi$ and $\xi_i$.\footnote{For RG flows in which the symmetries are broken explicitly, the scalars $\phi$ and $\xi_i$ should be thought of as compensator fields. For a discussion on compensator fields, see \cite{Komargodski:2011vj}.} We derive the effective action $S_{\text{eff}}[\phi,\xi_i]$ of $\phi$ and $\xi_i$ from basic symmetry principles. In particular, we show that  the effective action, up to four-fields four-derivative terms, can be written in the form\footnote{The explicit form of the effective action is given by equation (\ref{Gaction}).}
\be\label{intro:action}
S_{\text{eff}}[\phi,\xi_i]=S_{\rm conformal}[\phi; \Delta a]+S_{\rm global}[\xi_i; B_{ijkl}]+S_{\rm mixed}[\phi,\xi_i;\Delta a, b_i]\ .
\ee
The first term $S_{\rm conformal}[\phi; \Delta a]$ results from the conformal symmetry breaking alone and hence it is precisely the dilaton effective action of \cite{Komargodski:2011vj}. Similarly, $S_{\rm global}[\xi_i; B_{ijkl}]$ with coupling constants $B_{ijkl}$ is the part of the effective action that depends only  on the global symmetry breaking. Dimensionless coupling $B\equiv \{B_{ijkl}\}$ is a strongly paired symmetric 4-tensor which has the symmetries of the $N$-dimensional {\it elasticity tensor}. Finally, the mixed part of the action $S_{\rm mixed}[\phi,\xi_i;\Delta a, b_i]$ represents interactions between $\phi$ and $\xi_i$ with coupling constants $\Delta a$ and $b_i, i\in\{1,\cdots, N\}$. Interestingly, a part of $S_{\rm mixed}[\phi,\xi_i;\Delta a, b_i]$ is also universal. In general, $b_i$ and $B_{ijkl}$ depend on $\CFTUV$, $\CFTIR$, and deformations (or VEVs) that break the conformal symmetry in the UV. For unitary RG flows, these coefficients must also satisfy various positivity conditions which we will derive in this paper. 

There is a discrete difference between RG flows with and without global symmetry breaking. Nevertheless, the decomposition (\ref{intro:action}) of the effective action states that RG flows that do not break any global symmetries are a special case of  the general scenario with $\xi_i \rightarrow 0$ implying $S_{\text{eff}}[\phi,\xi_i]$ has a smooth $\xi_i\rightarrow 0$ limit. This in turn implies that breaking of additional global symmetries does not interfere with the proof of the 4d $a$-theorem by Komargodski and Schwimmer. This was already noticed by Bobev, Elvang, and Olson in \cite{Bobev:2013vta} for 4d RG flows with $U(1)$ symmetry breaking.

It is a fact that  scalar effective field theories in AdS$_d$ are in one-to-one correspondence with perturbative solutions of crossing symmetry in CFT$_{d-1}$ \cite{Heemskerk:2009pn,Heemskerk:2010ty,Fitzpatrick:2010zm,Penedones:2010ue,ElShowk:2011ag,Fitzpatrick:2011ia,Fitzpatrick:2011hu,Fitzpatrick:2011dm,Fitzpatrick:2012cg,Goncalves:2014rfa,Alday:2014tsa,Hijano:2015zsa,Aharony:2016dwx}. This connection was utilized in \cite{Kundu:2019zsl} to argue that every RG flow connecting two conformal fixed points in $d$ dimensions is equivalently described as  deformations of the spectrum of a generalized free CFT$_{d-1}$ for $d\ge 3$. In this paper, we adopt the same philosophy and analyze the dual CFT$_3$ description of the effective action (\ref{intro:action}) of NG bosons. The dual CFT$_3$ is obtained by deforming specific operator dimensions and OPE coefficients of a generalized free theory of $(N+1)$ scalar primary single-trace operators that are dual to NG bosons $\phi$ and $\xi_i$. This dual CFT$_{3}$, for any unitary RG flow, must obey the Euclidean axioms. This immediately implies that the space of $\{\Delta a,b_i ,B_{ijkl}\}$ for unitary RG flows can be constrained by invoking the chaos bound \cite{Maldacena:2015waa,Afkhami-Jeddi:2016ntf,Kundu:2020gkz} in the dual CFT$_3$. In particular, we argue that couplings $\Delta a,b_i ,$ and $B_{ijkl}$ must be positive definite.\footnote{To be specific, by  positive definiteness of the 4-tensor $B$ we simply mean that $B$ has a positive definite bi-quadratic form $B_{ijkl}c_i\c_jc_k\c_l > 0$ for all non-zero $c,\c\in \mathbb{R}^N$. This can be alternatively stated as $B$ is {\it strongly elliptic}. Note that there can be loop effects when $G$ is non-abelian, as we will explain later. Of course, $B$ should be understood as the 1-loop effective $B$ when loop effects are present.} Moreover, interference effects in the chaos bound impose further nonlinear constraints among $\{\Delta a,b_i ,B_{ijkl}\}$. These nonlinear analytic constraints, among other things, provide an upper bound on $\Delta a$ in terms of $b_i$ and $B_{ijkl}$.

\begin{figure}
\centering
\includegraphics[scale=0.33]{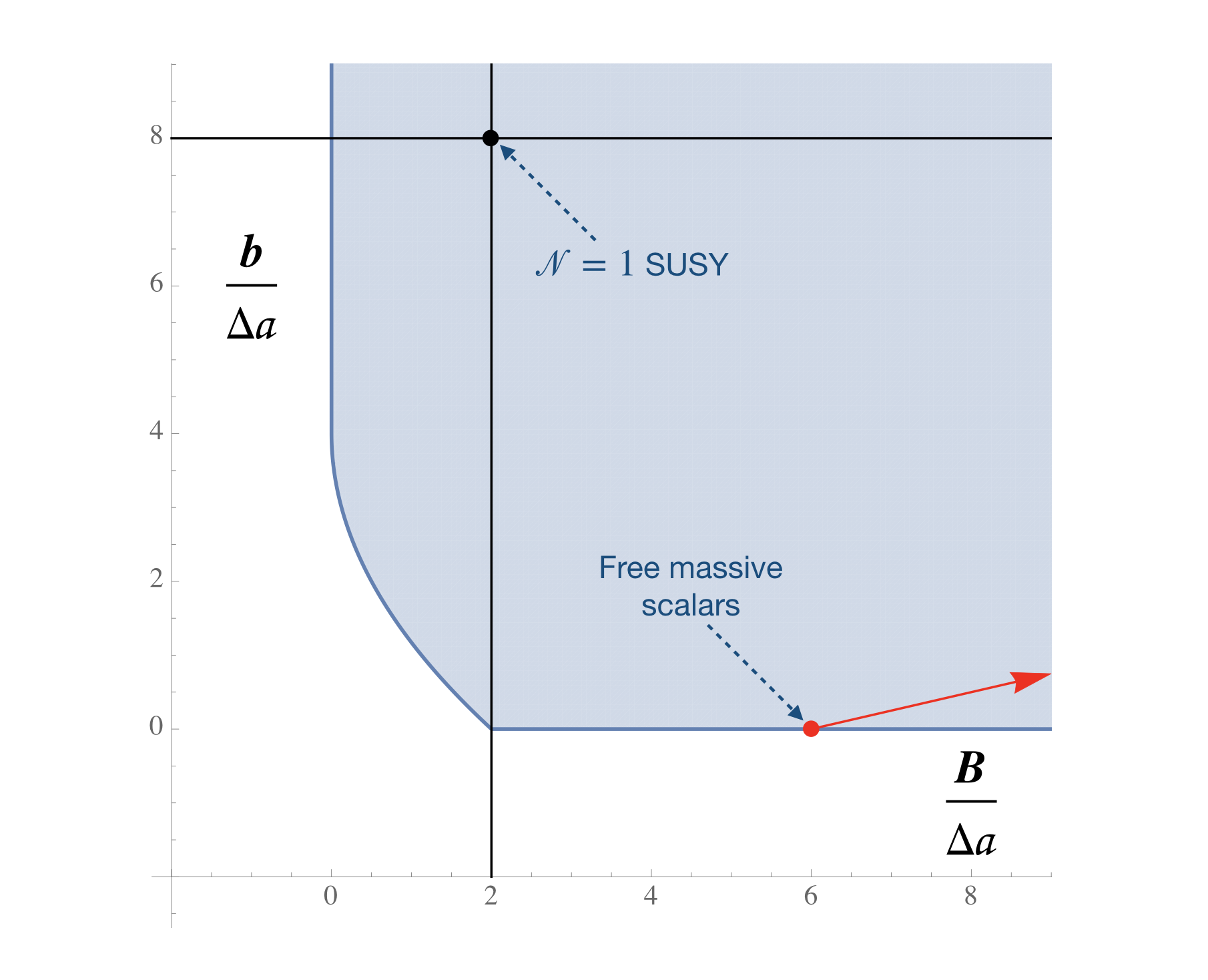}
\be
\Delta a=a_{\rm UV}-a_{\rm IR}> 0 \nonumber
\ee
\caption{ \label{intro:chaos} \small For unitary RG flows between two conformal fixed points in 4d with a broken $U(1)$ global symmetry only the shaded region (along with $\Delta a> 0$) is consistent with the chaos bound. Notice that there is a {\it kink} at $b=0$ and $B=2\Delta a$. The black dot corresponds to RG flows between two 4d $\N=1$ SCFTs in which  flows preserve  the $\N=1$ supersymmetry. The red dot corresponds to an RG flow in which the $\CFTUV$ is a theory of two massless scalars. This theory is deformed with mass terms that are infinitesimally different. The red line represents the same RG flow as we increase the mass difference. }
\end{figure}

As a representative example, we analyze RG flows between two conformal fixed points in 4d with a broken $U(1)$ global symmetry. Every such RG flow can be described in terms of exactly three parameters $\{\Delta a,b ,B\}$ that uniquely determine the low-energy effective action of NG bosons (see equation (\ref{action})) of  broken conformal and $U(1)$ symmetries. These RG flows, as we described before, have a dual description in terms of spectral deformations of a generalized free CFT in 3d of two scalar primary single-trace operators. For unitary RG flows, we invoke the chaos bound to constrain the space of $\{\Delta a,b,B\}$, as shown in figure \ref{intro:chaos}. Interestingly, there is a  ``bootstrap" kink in the exclusion plot \ref{intro:chaos}. However, we are not aware of any RG flow that sits on the kink.

In the exclusion plot \ref{intro:chaos}, we identify a special point in the allowed parameter space that corresponds to RG flows between two 4d $\N=1$ SCFTs in which the $\N=1$ supersymmetry is preserved along the flow. These flows break the $U(1)$ R-symmetry of the $\CFTUV$ since the stress tensor  and the R-current are in the same supermultiplet \cite{Bobev:2013vta, Schwimmer:2010za}. In the dual CFT$_3$ language, these supersymmetric flows are equivalently described by spectral deformations in which anomalous dimensions of certain double-trace  operators obey simple relations (see equation (\ref{bound:susy})).  Moreover, for $\N=1$ supersymmetric flows we show that there are infinitely many distinct $C$-functions that decrease monotonically from $a_{\rm UV}$ to $a_{\rm IR}$ under RG flows.

It should be noted that the same bounds can be obtained directly in flat space, perhaps with some additional assumptions about the scattering amplitudes.\footnote{For similar bounds on effective actions from scattering amplitudes see \cite{Remmen:2019cyz,Bellazzini:2020cot,Tolley:2020gtv,Trott:2020ebl}.} However, the dual CFT$_3$ description has several conceptual as well as technical advantages. For example, our derivation of the bounds does not make any assumptions about the dual CFT$_3$ beyond the usual Euclidean axioms. This simply means that some properties of low energy effective actions, such as (\ref{intro:action}), are more transparent in AdS. This parallels the idea of S-matrix bootstrap where conformal bootstrap methods are used to constrain QFTs in AdS \cite{Paulos:2016fap,Paulos:2016but,Paulos:2017fhb,Homrich:2019cbt}.

\subsection{RG and Quantum Chaos}
The CFT$_3$ based description of 4d RG flows naturally provides a set of 4d $C$-functions. Furthermore, this construction has the conceptual advantage that it explicitly relates $C$-functions with certain out-of-time-order correlators (OTOCs) in 3d which  have been used extensively in recent years as a  quantum field theoretic measure for chaos \cite{1969JETP...28.1200L,kitaev2014chaos,Maldacena:2015waa}. In particular, we will construct a series of $C$-functions all of which has the form  
\be\label{intro:C}
C(\mu) \sim \frac{1}{\beta} \lim_{t_*\gg t_0 \gg \beta} \int_{t_0}^{t_0+ \mu} dt_R\ \ e^{-2\pi t_R/\beta}\ \mbox{Re}\(F_d-F\(t_R-\frac{i \beta}{4}\)\)\ ,
\ee
where, $F_d$ and $F(t)$, as defined in \cite{Maldacena:2015waa}, are standard thermal correlators of simple operators that diagnose chaos.\footnote{Note that $C(\mu)$ in equation (\ref{intro:C}) is independent of $t_0$ as long as $t_0$ is much smaller than the effective scrambling time $t_*$. } To be specific, in the above expression $F(t_R-i\beta/4)$ is a CFT$_3$ four-point correlator of scalar primaries dual to NG bosons in the Minkowski vacuum state -- which we interpret as an OTOC in  a thermal state of temperature $1/\beta$ on Rindler space, where $t_R$ is the Rindler time. Monotonicity and positivity of  $C(\mu)$ follow directly from positivity conditions that  $F(t)$ satisfy \cite{Maldacena:2015waa} (see also \cite{Hartman:2016lgu,Afkhami-Jeddi:2016ntf,Kundu:2020gkz}). A special case of (\ref{intro:C}) is a set of  monotonically decreasing independent functions, also known as $a$-functions, that interpolate between $a_{\rm UV}$ in the UV ($\mu \rightarrow \infty$) and $a_{\rm IR}$ in the IR ($\mu \rightarrow 0$) establishing the RG irreversibility. As a byproduct, we obtain a relation between 4d $\Delta a$ and 3d OTOC
\be
\Delta a \propto \frac{1}{\beta} \lim_{t_*\gg t_0 \gg \beta} \int_{t_0}^{\infty} dt_R\ \ e^{-2\pi t_R/\beta}\ \mbox{Re}\(F_d-F\(t_R-\frac{i \beta}{4}\)\) > 0\ .
\ee

It is only natural to wonder whether there is a deeper, more fundamental connection between RG and chaos in QFT. At first sight, a more general connection seems unlikely. After all, chaos probes long-time but not necessarily low  energy properties of quantum systems.  So, it is not expected that the full richness of physics associated with chaos can be captured by RG which only deals with low energy degrees of freedom. However, information about the high energy degrees of freedom is not completely lost in any unitary RG flow. They are simply  hidden in the positivity and monotonicity properties of RG. The relation (\ref{intro:C}) connects these general features of RG with analogous monotonicity and positivity  properties of semiclassical chaos, however, in different spacetime dimensions. It is certainly possible that this connection is more fundamental and  holds even in the same spacetime dimensions.\footnote{A related but somewhat different question is how chaotic dynamics in QFT changes under RG flows. This has been discussed recently in \cite{Kundu:2020nir} for holographic theories.}

One significant hint for this general connection is that both RG and chaos are intimately related to causality.  This is certainly true in 4d  in which the $a$-theorem of \cite{Komargodski:2011vj} could be derived by invoking causality \cite{Adams:2006sv}. Moreover, for holographic theories, the RG monotonicity  follows directly from causality in general spacetime dimension \cite{Myers:2010xs,Myers:2010tj}. Likewise, the chaos bound of \cite{Maldacena:2015waa} is known to be related to causality as well \cite{Hartman:2016lgu,Afkhami-Jeddi:2016ntf,Kundu:2020gkz}. 

There is another nice interrelation between RG, chaos, and causality in 4d CFT.  Any unitary CFT must obey certain causality constraints that are known as the conformal collider bounds \cite{Hofman:2008ar,Komargodski:2016gci,Hartman:2016dxc, Hofman:2016awc}. The collider bounds can also be thought of as a special  case of the chaos bound for vacuum CFT correlators \cite{Hartman:2016lgu}. In 4d, the collider bounds impose that the Euler central charge $a$ must be positive. This positivity together with the 4d $a$-theorem then imply that   the Euler central charge is a good measure of the effective number of degrees of freedom in 4d CFT.

All these hints are suggestive of a much deeper relationship between RG and chaos.  It would be very interesting to make this connection more direct and explicit. For example, chaos in QFT could be formulated as coarse-graining of the operator algebra. Such a description of chaos does exist in quantum mechanics \cite{Radicevic:2016kpf}. It is also possible that both RG and semiclassical chaos are related by some version of the eigenstate thermalization hypothesis (ETH) \cite{Srednicki:1995pt,PhysRevA.43.2046,2008Natur.452..854R}.

\subsection{Outline}
The rest of the paper is organized as follows. We begin with a detailed analysis of 4d RG flows with a broken $U(1)$ global symmetry in section \ref{sec:u1}. In section \ref{sec:susy}  we discuss 4d  RG flows with $\N=1$ supersymmetry and compare it with our general results. In section \ref{sec:general} we derive the most general low energy effective action that describes 4d RG flows with a broken continuous symmetry group $G$, where  $G$ can be a direct product of finite number of simple Lie groups. Furthermore, we derive constraints on this effective action for unitary RG flows. In section \ref{sec:C} we construct $C$-functions that have the form (\ref{intro:C}). Finally, in section \ref{sec:scalars} we provide a simple example which highlights basic features of our general framework. We take the $\CFTUV$ to be a theory of two massless scalars. The conformal symmetry and the global $U(1)$ symmetry are broken explicitly by introducing different mass terms for two scalars. The $\CFTIR$ is this case is trivial with no degrees of freedom.

\section{RG Flows with a Broken $U(1)$ Global Symmetry}\label{sec:u1}
In this section we consider 4d CFTs with some $U(1)$ global symmetry  in which the conformal symmetry and the $U(1)$ symmetry are broken either spontaneously or explicitly. We assume that the induced flow terminates in a different CFT  in the deep IR which is invariant under the same $U(1)$ transformation. Every such RG flow can be described as spontaneous breaking of conformal and $U(1)$ symmetries. This enables us to study general features of these RG flows in terms of the effective action of some Nambu-Goldstone bosons of  spontaneously broken conformal and $U(1)$ symmetries.

\subsection{The Dilaton-Axion Effective Action}
Consider a $\CFTUV$ in (3+1)-dimensions with a global $U(1)$ symmetry. We assume that the $\CFTUV$ has a moduli space of vacua which enables us to break the conformal symmetry and the $U(1)$ symmetry spontaneously by turning on VEVs for an operator $O$. The VEV $\langle O\rangle\sim f$ triggers an RG flow that leads to some $\CFTIR$ which we assume to be invariant under the UV $U(1)$ symmetry.\footnote{It should be noted that there could be other emergent $U(1)$ symmetries in the IR that do not embed at all in the UV theory. These additional $U(1)$s will not affect our argument. } In other words, the global $U(1)$ symmetry  of the UV theory is also a symmetry of $\CFTIR$ (which can  be anomalous in the presence of background fields). Of Course, the $\CFTIR$ can transform trivially under the UV $U(1)$ symmetry group.

Each broken generator associated with spontaneous breaking of continuous global symmetries produces a  massless Nambu-Goldstone (NG) {\it pseudo-scalar}. The low energy effective action of the NG bosons can be obtained in a systematic way by using the coset construction introduced in \cite{Coleman:1969sm,Callan:1969sn} (see also \cite{Weinberg:1996kr}). The coset construction for spontaneous breaking of space-time symmetries is more subtle \cite{Volkov:1973vd}. When the conformal algebra is spontaneously broken to Poincaré sub-algebra
\be
\mathfrak{so}(4,2)\rightarrow \mathfrak{iso}(3,1)\ ,
\ee
one may expect that there are five NG modes -- a scalar $\tau$ associated with the broken dilation generator $D$ and a vector $a_\mu$ associated with the broken special conformal generators $K_\mu$. However, not all these modes are independent because of the inverse Higgs effect \cite{Ivanov:1975zq}. This follows from the fact that the commutator $[K_\mu, P_\nu]=2\(J_{\mu\nu}-\eta_{\mu\nu}D\)$ can be utilized to eliminate $a_\mu=\frac{1}{2} \p_\mu e^\tau$ \cite{Salam:1970qk,Isham:1970xz,Isham:1970gz,Low:2001bw,McArthur:2010zm,Hinterbichler:2012mv}.

So, the spontaneously broken conformal symmetry generates only one massless NG boson -- the dilaton $\tau$.  The dilaton is accompanied by a pseudo-scalar $\beta$, which is the NG boson of  the spontaneously broken $U(1)$ symmetry. For RG flows in which the conformal symmetry and the $U(1)$ symmetry are broken explicitly, the dilaton $\tau$ and the axion $\beta$ can be introduced as compensators for broken symmetries.  So, in general the low energy theory consists of $\CFTIR$ and massless scalars $\tau$ and $\beta$
 \be\label{IR}
 S_{\rm IR}=\CFTIR+ S_{\rm eff}[\tau,\beta]\ .
 \ee
The effective action $S_{\rm eff}[\tau,\beta]$ can be obtained by using the coset construction. However, following \cite{Bobev:2013vta} we will derive the effective action in a physically more transparent way  by coupling the theory to background fields. 

We begin by coupling the theory to a background metric $g_{\mu\nu}(x)$ and a background $U(1)$ gauge potential $A_\mu(x)$. In the presence of background fields, the conformal trace anomaly has the following structure 
 \be\label{anomaly0}
 \langle T^\mu_\mu\rangle =- a E_4+ c W^2+\kappa_0 F^2\ ,
 \ee
up to total derivative terms which can be removed by adding finite and covariant counter-terms in the UV theory. Here, $E_4$ is the 4d Euler density, $W_{\mu\nu\alpha\beta}$ is the Weyl tensor and $F=dA$ is the flux for the background gauge field. Global symmetries can also have 't Hooft anomalies.
In 4d, such anomalies reveal themselves  through the current $j_\mu$ associated with the $U(1)$ symmetry which is no longer conserved 
\be
 \langle \nabla_\mu j^\mu\rangle = c_1 F_{\mu\nu}\tilde{F}^{\mu\nu}+ c_2 R_{\mu\nu\alpha\beta}\tilde{R}^{\mu\nu\alpha\beta}\ .
\ee
Note that Hodge dualization
\be
\tilde{F}_{\mu\nu}=\frac{1}{2}\epsilon_{\mu\nu\alpha\beta} F^{\alpha\beta}\ , \qquad \tilde{R}_{\mu\nu\alpha\beta}=\frac{1}{2}\epsilon_{\mu\nu\gamma\delta}R^{\gamma\delta}_{\ \ \alpha\beta}
\ee
are defined with respect to the background metric $g_{\mu\nu}$. Since, the global symmetry is anomalous, one may worry that the trace anomaly \eqref{anomaly0} can also have non-gauge invariant terms. However, as shown in \cite{Bobev:2013vta}, the Wess-Zumino consistency conditions guarantee that the trace anomaly is gauge-invariant. 

In the IR, the gauge field $A_\mu$ may not couple to $\CFTIR$ at all or it can couple to some spin-1 abelian conserved current $j_\mu^{\rm IR}$ of $\CFTIR$. In the latter case, the $U(1)$ symmetry associated with $j_\mu^{\rm IR}$ can also have 't Hooft anomalies.  The  standard anomaly matching arguments of \cite{Schwimmer:2010za} imply that the IR theory (\ref{IR}) must have the same anomalies as the UV theory $\CFTUV$. This requirement completely fixes the low energy effective action $S_{\rm eff}[g_{\mu\nu},A_\mu;\tau,\beta]$. The flat space limit  of $S_{\rm eff}[g_{\mu\nu},A_\mu;\tau,\beta]$ with no background gauge field then leads to the effective action $S_{\rm eff}[\tau,\beta]$.

Let us now study the variation of the action (\ref{IR}) under diff$\times$Weyl transformations and gauge transformations.  Under Weyl transformations 
 \be\label{weyl}
 g_{\mu\nu}(x)\rightarrow e^{2\sigma(x)}g_{\mu\nu}(x)\ , \qquad \tau(x)\rightarrow \tau(x)+\sigma(x)\ .
 \ee
Similarly, the gauge transformation is defined in the usual way 
\be\label{gauge}
A_\mu(x) \rightarrow A_\mu(x) +\nabla_\mu \alpha(x)\ , \qquad \beta(x)\rightarrow\beta(x)+ \alpha(x)\ .
\ee
Of course, in general $\CFTUV$ and $\CFTIR$ have different set of anomalies. Hence, all changes in anomalies  in the flow from  $\CFTUV$ to $\CFTIR$ must be compensated by the dilaton and the axion. Hence, the Weyl variation of the effective action is completely fixed
\be\label{variation1}
\delta_\sigma S_{\rm eff}[g_{\mu\nu},A_\mu;\tau,\beta]=\int d^4x \sqrt{-g} \sigma(x)\left(-\Delta a E_4+  \Delta c W^2+\Delta \kappa_0 F^2\right)\ .
\ee
Likewise, variation of the effective action under the gauge transformation (\ref{gauge}) is also fixed
\be\label{variation2}
\delta_\alpha S_{\rm eff}[g_{\mu\nu},A_\mu;\tau,\beta]=\int d^4x \sqrt{-g} \alpha(x)\left(\Delta c_1 F_{\mu\nu}\tilde{F}^{\mu\nu}+\Delta c_2 R_{\mu\nu\alpha\beta}\tilde{R}^{\mu\nu\alpha\beta}\right)\ .
\ee
In the above equations $\Delta (\cdots)$ denotes the change of an anomaly under the RG flow, where IR anomalies should be understood as the total anomalies of $\CFTIR$, the  dilaton, and the axion. The variational equations (\ref{variation1}) and (\ref{variation2}) can now be solved systematically to obtain $S_{\rm eff}[g_{\mu\nu},A_\mu;\tau,\beta]$ by a straightforward generalization of \cite{Bobev:2013vta}. 

It is useful, as discussed in \cite{Kundu:2019zsl}, to decompose the effective action in the following way
\begin{align}\label{simple0}
S_{\rm eff}[g_{\mu\nu},A_\mu;\tau,\beta]&=\int d^4x \sqrt{-g} \beta(x)\left(\Delta c_1 F_{\mu\nu}\tilde{F}^{\mu\nu}+\Delta c_2 R_{\mu\nu\alpha\beta}\tilde{R}^{\mu\nu\alpha\beta}\right)\nonumber\\
&+\int d^4x \sqrt{-g} \tau(x)\left(-\Delta a E_4+  \Delta c W^2+\Delta \kappa_0 F^2\right)+\delta S_{\rm WZ}+S_{\rm inv}\ .
\end{align}
Note that the first term in the above equation generates the correct gauge variation (\ref{variation2}).  On the other hand, the second term in the above equation  generates the correct Weyl variation (\ref{variation1}) plus an extra term $-\Delta a\int d^4x \sqrt{-g} \tau(x)\delta_\sigma E_4$ which is cancelled by adding  a non-linear Wess-Zumino term $\delta S_{\rm WZ}$ of $\tau$. Of course, we can also add a term $S_{\rm inv}$ in the action whose gauge and Weyl variations vanish. The main advantage of this formalism is that $\delta S_{\rm WZ}$ is uniquely fixed by $\Delta a$ \cite{Komargodski:2011vj}
\be
\delta S_{\rm WZ}=- \Delta a\int d^4x \sqrt{-g}\(4\(R^{\mu\nu}-\frac{1}{2} g^{\mu\nu}R\)\nabla_\mu\tau \nabla_\nu \tau -2\(\nabla \tau\)^2\(2\Box \tau-\(\nabla \tau\)^2\)\)
\ee
up to terms that are invariant under both diff$\times$Weyl  transformations and gauge transformations and hence can be absorbed in $S_{\rm inv}$. 

Importantly, only $\delta S_{\rm WZ}$ and $S_{\rm inv}$ in (\ref{simple0}) contribute in the flat space limit with no background gauge field. We now focus on $S_{\rm inv}$. This is the part of the effective action which, in general, depends on the details of the RG flow. Nevertheless, at each derivative order only a finite number of independent gauge and Weyl invariant terms can appear in  $S_{\rm inv}$.\footnote{The $S_{\rm inv}$ is constructed from terms that are exactly invariant under the gauge and Weyl transformations. Hence, it is possible that we miss Wess-Zumino type terms in the action that are not exactly invariant but shifts by a total derivative under  the gauge and Weyl transformations \cite{DHoker:1994rdl}. However, these terms do not contribute at the 4-field 4-derivative level and can be ignored for our purpose.} These terms can be efficiently constructed by defining gauge and Weyl invariant combinations
\be\label{metric0}
\hat{g}_{\mu\nu}=e^{-2\tau}g_{\mu\nu}\ , \qquad \hat{A}_\mu=A_\mu -\nabla_\mu \beta \ .
\ee
Up to four derivatives, the most general $S_{\rm inv}$ is given by \cite{Bobev:2013vta}
\begin{align}\label{inv}
S_{\rm inv}=\int d^4x \sqrt{-\hat{g}}\left(-\frac{f^2}{2}\( \frac{\hat{R}}{6}+\gamma_0^2 \hat{g}^{\mu\nu}\hat{A}_\mu\hat{A}_\nu\) +\sum_{i=1}^9 \gamma_i W_i+\O(\nabla^6)\right)\ ,
\end{align}
where $\hat{R}$, is defined using the Weyl-invariant metric (\ref{metric0}) and four derivative invariants $W_i$ are given in appendix \ref{app:U1}. Note that $f$ has dimension of mass and $\gamma_i$ are real dimensionless coefficients. We are now ready to write down the low-energy effective action by taking the flat space limit of (\ref{simple0}) with no background gauge field. Putting everything together, $S_{\text{eff}}[\tau,\beta]$ is given by
\begin{align}
S_{\text{eff}}[\tau,\beta]=& \int d^4x \(-\frac{f^2}{2}e^{-2\tau}\left(\(\p \tau\)^2+\gamma_0^2 \(\p \beta\)^2\)+2\Delta a \(\p \tau\)^2 \(2\Box \tau-\(\p \tau\)^2\)\)\nonumber \\
&+\int d^4x\ e^{-4\tau}\left(\sum_{i=1}^9 \gamma_i W_i\)_{g_{\mu\nu}=\eta_{\mu\nu}, A_\mu=0}+\cdots\ ,
\end{align} 
where dots represent higher derivative terms. Equations of motion at the two derivative level are given by
\be\label{eq:eom}
\Box \tau=\(\p \tau\)^2-\gamma_0^2 \(\p \beta\)^2\ , \qquad \Box \beta=2\(\p \tau \cdot \p \beta\)\ .
\ee
Terms that vanish once we impose the on-shell condition for the dilaton and the axion can be safely ignored at low energies since these terms can only affect low energy observables at subleading orders. Hence, the above effective action can be further simplified by using the above equations of motion, yielding
\begin{align}\label{simple}
S_{\text{eff}}[\tau,\beta]=& \int d^4x \(-\frac{f^2}{2}e^{-2\tau}\left(\(\p \tau\)^2+ \(\p \beta\)^2\)+2\Delta a \(\p \tau\)^4 - 4\Delta a \(\p \tau\)^2 \(\p \beta\)^2\)\nonumber \\
&+\int d^4x \(B\(\p \beta\)^4+b \(\p \tau \cdot \p \beta\)^2\) +\cdots\ ,
\end{align}
where, we have redefined $\beta$ to absorb $\gamma_0$. Note that coefficients $b$ and $B$ are some linear combinations of $\gamma_i$ (see appendix \ref{app:U1}).

\subsection{Physical Dilaton and Axion}
The effective action (\ref{simple}) is not very useful when we wish to study the theory using traditional tools of QFT. We resolve this issue by a simple field redefinition: 
\be
e^{-(\tau+i \beta)}=1-\frac{\phi+i \xi}{f}\ , 
\ee
where, the {\it physical} fields $\phi$ and $\xi$ are real. Plugging this into the action (\ref{simple}) and then expanding up to fourth order in the fields, we obtain
\begin{align}\label{action}
S_{\text{eff}}[\phi,\xi]=\int d^4x & \(-\frac{1}{2}\(\p \phi\)^2-\frac{1}{2}\(\p \xi\)^2+\frac{\Delta a}{2f^4}\(\phi^2 \Box^2 \phi^2-2\xi^2 \Box^2 \phi^2\)\)\nonumber\\
&+\frac{1}{4f^4}\int d^4x\(B \xi^2 \Box^2 \xi^2+b \phi \xi \Box^2 \phi \xi\)+\cdots\ ,
\end{align}
where we have used the equations of motion to simplify the action. The first line of the action is completely fixed by the UV and the IR fixed points of the RG flow. On the other hand, the second line depends on the details of the RG flow and parameters $B$ and $b$, in general, are completely arbitrary. Dots represent terms with more than four fields and/or four derivatives.

To summarize, any 4d RG flow with  $U(1)$ global symmetry breaking between two CFTs can be described by the effective action of NG bosons of  spontaneously broken conformal and $U(1)$ symmetry. Up to four derivative order, the effective action is completely fixed in terms of three parameters $\{\Delta a, b, B\}$. Also the effective action (\ref{action}) has the structure (\ref{intro:action}) implying RG flows that do not break the global symmetry are a special case of  the general scenario with $\xi \rightarrow 0$. There is a discrete difference between RG flows with and without global symmetry breaking, however $S_{\text{eff}}[\phi,\xi]$ still has a smooth $\xi\rightarrow 0$ limit.

The above feature of the effective action (\ref{action}), as correctly pointed out in \cite{Bobev:2013vta}, has an important implication. The 4-particle interaction of the physical dilaton remains unmodified even when we break the global $U(1)$ symmetry. This implies that the proof of the $a$-theorem by Komargodski and Schwimmer applies here as well. Moreover, from the action (\ref{action}) it is clear that there are other constraints on the parameters $\{\Delta a, b, B\}$ for unitary RG flows. Next we will introduce an equivalent CFT$_3$ description of these RG flows to impose constraints on $\{\Delta a, b, B\}$ from the chaos bound.

\subsection{Dual CFT$_3$ Description}
It was shown in \cite{Kundu:2019zsl} that every RG flow connecting two conformal fixed points in $d$ dimensions can be interpreted as  deformations of the spectrum of a generalized free CFT$_{d-1}$ for $d\ge 3$. This dual CFT$_{d-1}$, for any unitary RG flow, must obey the Euclidean axioms. As a consequence, 4-point correlators of the dual CFT$_{d-1}$ must obey the chaos bound \cite{Maldacena:2015waa}. This imposes rigorous  constraints on $\{\Delta a, b, B\}$ for unitary RG flows.

We analyze the effective action (\ref{action}) as a theory in AdS$_4$ with AdS radius $R_{\text{AdS}}$ large but finite. The action now is simply given by
\be\label{ads}
S_{\text{eff}}[\phi,\xi]=\int d^4x \sqrt{g_{\rm AdS}}\left(-\frac{1}{2}g^{\mu\nu}_{AdS}\partial_\mu \phi \partial_\nu \phi-\frac{1}{2}g^{\mu\nu}_{AdS}\partial_\mu \xi \partial_\nu \xi+ \mathcal{L}_{\rm int}\right)\ ,
\ee
where the interactions are obtained from (\ref{action})
\be\label{int}
\mathcal{L}_{\rm int}=\frac{1}{4f^4}\(2\Delta a\phi^2 \Box^2 \phi^2-4\Delta a\xi^2 \Box^2 \phi^2+B \xi^2 \Box^2 \xi^2+b \phi \xi \Box^2 \phi \xi\)+\cdots\ .
\ee
This is a theory in AdS without dynamical gravity. In the dual CFT$_3$, the  stress tensor decouples from the low energy spectrum. In other words, the CFT$_3$ central charge $c_T\rightarrow \infty$, however,  $f R_{\text{AdS}}\equiv \Delta_{f}$ is large but fixed.\footnote{The central charge $c_T$ is the overall coefficient that appears in the stress tensor two-point function.} The resulting CFT$_3$ must be well behaved below the cut-off scale $\Delta_f$.  This {\it effective} CFT$_3$ contains two scalar primary operators $\O_\phi$ and $\O_\xi$ which are dual to the dilaton and the axion respectively. The fact that $\phi$ and $\xi$ are NG bosons implies that 
\be
\Delta_\phi=\Delta_\xi=3\ 
\ee
will not receive perturbative corrections. 

We follow the formalism developed in \cite{Kundu:2019zsl} and interpret the dual CFT$_3$ as a small perturbation of a generalized free CFT in 3d with two scalar primaries. First consider the case, $\mathcal{L}_{\rm int}=0$.  The dual CFT$_3$, in this case, is exactly a generalized free CFT of scalar primaries $\O_\phi$ and $\O_\xi$.  In addition, crossing symmetry requires that this generalized free CFT must also contain infinite towers of double-trace operators $[\O_\phi \O_\phi]_{n,\ell}$, $[\O_\xi \O_\xi]_{n,\ell}$, and $[\O_\phi \O_\xi]_{n,\ell}$ with spin $\ell$ and dimension $6+2n+\ell$ for all integer $n\ge 0$ \cite{Komargodski:2012ek,Fitzpatrick:2012yx}. We now turn on the interaction $\mathcal{L}_{\rm int}$ in AdS$_4$. The bulk theory (\ref{ads}) now corresponds to a deformed solution of CFT$_3$ crossing equations in which double-trace operators $[\O_\phi \O_\phi]_{n,\ell}$, $[\O_\xi \O_\xi]_{n,\ell}$, and $[\O_\phi \O_\xi]_{n,\ell}$ acquire anomalous dimensions $\gamma_{n,\ell}^{(\phi\phi)}$, $\gamma_{n,\ell}^{(\xi\xi)}$, and $\gamma_{n,\ell}^{(\phi\xi)}$ respectively. The information of $\{\Delta a, b, B\}$ is contained in these anomalous dimensions.

\subsection{CFT Regge Correlators}\label{sec:regge}
We are now in a position to study Lorentzian four-point functions for the CFT$_3$ dual to the effective field theory (\ref{ads}). First, we start with two-point functions which can be easily computed from (\ref{ads})
\be
\langle \O_\phi(x_1) \O_\phi(x_2)\rangle=\langle \O_\xi(x_1) \O_\xi(x_2)\rangle  =\frac{c_0}{|x_1-x_2|^{6}}\ ,
\ee
 where $c_0=\frac{12}{\pi^2}$.

\begin{figure}
\centering
\includegraphics[scale=0.5]{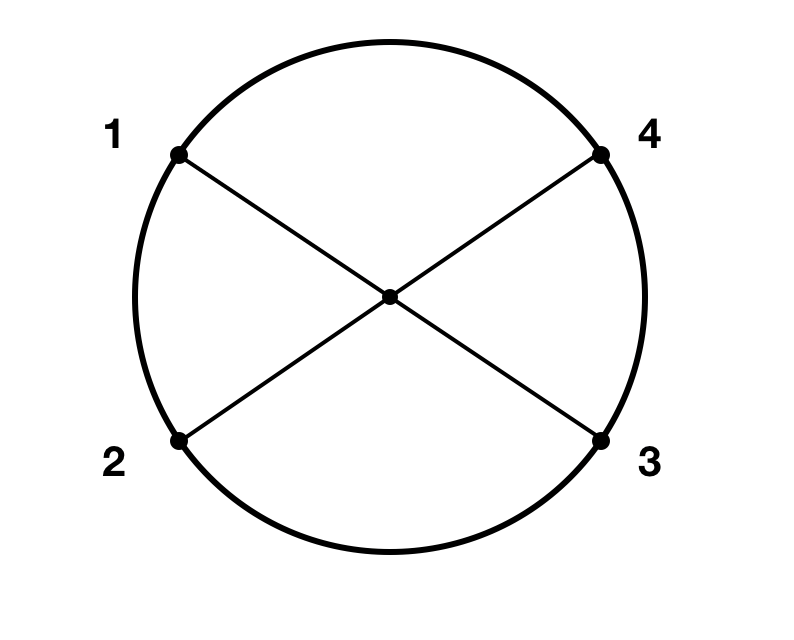}
\caption{ \label{fig:witten} \small  The tree-level contact Witten diagram.}
\end{figure}

Next we consider various four-point functions of operators $\O_\phi$ and $\O_\xi$. We are interested in the contributions of the four-point interaction $\mathcal{L}_{\rm int}$ in the bulk theory (\ref{ads}) to these four-point correlators. These are obtained from  the tree-level Witten diagram in figure \ref{fig:witten}.
We begin with the Lorentzian correlator $G_{\phi\phi\phi\phi}(\r,\br)=\langle \O_\phi(x_4) \O_\phi(x_1) \O_\phi(x_2) \O_\phi(x_3)  \rangle$ where all points are restricted to a $2$d subspace:
\begin{align}\label{points}
x_1=-x_2=(x^-=\rho,x^+=-\bar{\rho},0)\ ,  \qquad x_3=-x_4=(x^-=-1,x^+=1,0)\ ,
\end{align}
with $0<\rho<1$ and $\br>1$. Note that we are using null coordinates $x^{\pm}=x^0\pm x^1$, where $x^0$ is time (see figure \ref{config}). The CFT Regge limit is defined as
\be\label{regge2}
\rho\rightarrow \infty\ ,\quad \br\rightarrow 0\ , \qquad \text{with} \qquad \r \br=\text{fixed}>0
\ee
of the Lorentzian correlator $\langle \O_\phi(x_4) \O_\phi(x_1) \O_\phi(x_2) \O_\phi(x_3)  \rangle$, where operators are ordered as written. Our goal is to compute the contribution of $\mathcal{L}_{\rm int}$ to $G_{\phi\phi\phi\phi}(\r,\br)$ in the Regge limit (\ref{regge2}). We follow \cite{Kundu:2019zsl}  to obtain the leading Regge contribution
\be
G_{\phi\phi\phi\phi}(\r,\br)\approx \frac{c_0^2}{(16\r\br)^3}+ i \frac{\Delta a}{16 \pi^5 \Delta_f^4}\frac{\rho}{(\r\br)^{7/2}} f_{3333}\(-\frac{1}{2}\log (\r\br)\)
\ee
where the first term comes from the bulk identity exchange (disconnected Witten diagram). The function $f_{3333}$ is an integral 
\begin{align}\label{def:f}
f_{3333}(s)=\int_{-\infty}^\infty d\nu \Omega_{i\nu}(s)\Gamma\left(\frac{13/2+i\nu}{2} \right)^2 \Gamma\left(\frac{13/2-i\nu}{2} \right)^2
\end{align}
of the Harmonic function $\Omega_{i\nu}$ on hyperbolic space $H_{2}$.\footnote{ Harmonic functions $\Omega_{i\nu}$  are known in any dimension \cite{Costa:2017twz}.} The exact expression for $f_{3333}(s)$ will not be important for us. The only relevant information is that  $f_{3333}\(-\frac{1}{2}\log (\r\br)\)>0$ for $0<\r\br<1$. 

Similarly,  the Lorentzian correlator $G_{\xi\xi\xi\xi}(\r,\br)=\langle \O_\xi(x_4) \O_\xi(x_1) \O_\xi(x_2) \O_\xi(x_3)  \rangle$ can be computed in an identical way. In particular,   the leading Regge contribution is given by
\be
G_{\xi\xi\xi\xi}(\r,\br)\approx \frac{c_0^2}{(16\r\br)^3}+ i \frac{B}{32\pi^5  \Delta_f^4}\frac{\rho}{(\r\br)^{7/2}} f_{3333}\(-\frac{1}{2}\log (\r\br)\)\ .
\ee

There are many mixed  four-point functions that we can construct with operators $\O_\phi$ and $\O_\xi$. Clearly, the four-point interaction $\mathcal{L}_{\rm int}$ can only contribute to mixed correlators with two  $\O_\phi$ operators and two $\O_\xi$ operators. For example, consider the correlator 
\be
G_{\phi\phi\xi\xi}(\r,\br)\equiv\langle \O_\phi(x_4) \O_\xi(x_1) \O_\xi(x_2) \O_\phi(x_3)  \rangle=\langle \O_\xi(x_4) \O_\phi(x_1) \O_\phi(x_2) \O_\xi(x_3)  \rangle\ .
\ee
At the leading order in the Regge limit, by following \cite{Kundu:2019zsl}, we obtain
\be
G_{\phi\phi\xi\xi}(\r,\br)\approx \frac{c_0^2}{(16\r\br)^3}+ i \frac{b}{128 \pi^5  \Delta_f^4}\frac{\rho}{(\r\br)^{7/2}} f_{3333}\(-\frac{1}{2}\log (\r\br)\)\ .
\ee
Note that the same $f_{3333}$ function appears here as well. 

Moreover, the correlator $G_{\phi\xi\phi\xi}(\r,\br)\equiv \langle \O_\phi(x_4) \O_\xi(x_1) \O_\phi(x_2) \O_\xi(x_3)  \rangle$ and its cousins also grow in the limit $\r\rightarrow \infty$ for fixed $\r\br$. In particular, at tree level all these correlators have the following Regge behavior
\be
G_{\phi\xi\phi\xi}(\r,\br)\approx  i \frac{\(b-8\Delta a\)}{256 \pi^5  \Delta_f^4}\frac{\rho}{(\r\br)^{7/2}} f_{3333}\(-\frac{1}{2}\log (\r\br)\)\ .
\ee

\subsection{Anomalous Dimensions}
The bulk theory (\ref{ads})  leads to anomalous dimensions to double-trace operators $[\O_\phi \O_\phi]_{n,\ell=2}$, $[\O_\xi \O_\xi]_{n,\ell=2}$, and $[\O_\phi \O_\xi]_{n,\ell=2}$.
Among these double twist operators, the operators $[\O_\phi \O_\phi]_{0,2}$, $[\O_\xi \O_\xi]_{0,2}$, and $[\O_\phi \O_\xi]_{0,2}$ are of particular importance. So, we introduce the notation
\be
\gamma_{\phi\phi}\equiv \gamma_{0,2}^{(\phi\phi)}\ , \qquad \gamma_{\xi\xi}\equiv \gamma_{0,2}^{(\xi\xi)}\ , \qquad \gamma_{\phi\xi}\equiv \gamma_{0,2}^{(\phi\xi)}
\ee
to denote anomalous dimensions of spin-2 double-trace operators with minimal twists.

From the Regge correlators of the previous section, we can  relate $\{\Delta a,b,B\}$ to anomalous dimensions $\gamma_{\phi\phi}$, $\gamma_{\xi\xi}$, and $\gamma_{\phi\xi}$. Following \cite{Kundu:2019zsl}, we find
\be\label{anomalous}
\gamma_{\phi\phi}=-\frac{704  \Delta a}{13\pi^2\Delta_f^4}\ , \qquad \gamma_{\xi\xi}=-\frac{352 B}{13  \pi^2\Delta_f^4}\ , \qquad \gamma_{\phi\xi}=\frac{88\(8\Delta a-b\)}{13  \pi^2 \Delta_f^4}\ .
\ee
There are general constraints on families of minimal twist operators that appear in the OPEs of primary operators of any unitary CFTs in more than two dimensions \cite{Komargodski:2012ek,Kundu:2020gkz}. It is tempting to apply these constraints directly to \eqref{anomalous}, however one should be more careful for the following reason. The dual CFT$_3$ must be regarded as an effective CFT which is defined order by order in perturbation theory. Of course, even for such a theory bounds of \cite{Komargodski:2012ek,Kundu:2020gkz} apply to minimal twist operators. However, identifying  families of minimal twist operators can be subtle  for an effective CFT. In particular, it is easy to obtain a wrong bound when the anomalous dimension and the OPE coefficient of a candidate minimal twist operator receive contributions at different orders in perturbation theory.\footnote{For example, consider the stress tensor operator which has twist 1. Obviously, it appears in the OPE of $\O_\phi\O_\phi$, as well as $\O_\xi \O_\xi$. Hence, the stress tensor is truly the lowest twist spin-2 operator in the full theory. However, in the limit $c_T \rightarrow \infty$, the stress tensor contribution to 4-point correlators is subleading. So, it is unclear whether, and in what sense, the CFT Nachtmann theorems of \cite{Komargodski:2012ek,Kundu:2020gkz} apply to \eqref{anomalous}. } Therefore, we will not apply   the CFT Nachtmann theorem directly to \eqref{anomalous}. Instead, we will utilize the chaos bound which leads to similar but not exactly equivalent constraints. Positivity conditions obtained from the chaos bound are more reliable since they follow from rigorous CFT sum-rules \cite{Afkhami-Jeddi:2016ntf,Afkhami-Jeddi:2017rmx,Afkhami-Jeddi:2018own}.

\subsection{Constraints from the Chaos Bound}\label{sec:otoc}

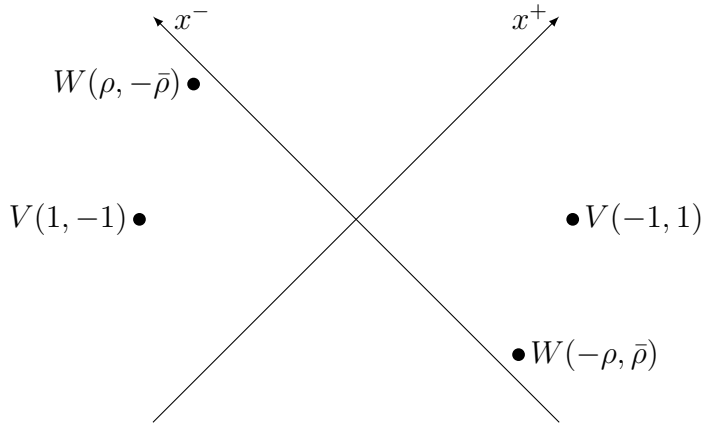
\begin{figure}
\begin{center}
\begin{tikzpicture}[baseline=-3pt, scale=1.80]
\begin{scope}[very thick,shift={(4,0)}]
\coordinate (v1) at (-1.5,-1.5) {};
\coordinate(v2) at (1.5,1.5) {};
\coordinate (v3) at (1.5,-1.5) {};
\coordinate(v4) at (-1.5,1.5) {};
\draw[thin,-latex]  (v1) -- (v2)node[left]{$x^+$};
\draw[thin,-latex]  (v3) -- (v4)node[right]{$\ x^-$};
\draw(-1.6,0)node[left]{ $\ V(1,-1)$};
\draw(1.6,0)node[right]{ $ V(-1,1)$};
\filldraw[black]  (-1.6,0) circle (1pt);
\filldraw[black]  (1.6,0) circle (1pt);
\coordinate(v5) at (0,0) {};
\def \fac {.6};
\filldraw[black]  (-1.2,1) circle (1 pt);
\filldraw[black]  (1.2,-1) circle (1pt);
\draw(-1.2,1)node[left]{ $ W(\rho,-\bar{\rho})$};
\draw(1.2,-1)node[right]{ $ W(-\rho,\bar{\rho})$};
\end{scope}
\end{tikzpicture}
\end{center}
\caption{\label{config} \small A Lorentzian four-point function of $W=\O_\phi+c_1 \O_\xi$ and $V=\O_\phi+c_2 \O_\xi$. All points are restricted to a $2$d subspace $\{x^0,x^1\}$ and time $x^0$ is running upward. Null coordinates are defined as $x^{\pm}=x^0\pm x^1$. }
\end{figure}

We  now  impose constraints on the effective action (\ref{action}) by utilizing the chaos bound in the dual CFT$_3$. Consider the Lorentzian correlator 
\be\label{eq:correlator}
G=\frac{\langle V(x_4) W(x_1) W(x_2) V(x_3)  \rangle}{\langle  W(x_1) W(x_2)   \rangle\langle V(x_4) V(x_3)  \rangle}
\ee
in the Regge kinematics (\ref{points}), as shown in figure \ref{config}, where operators inside the correlator are ordered as written. In the above correlator, $W$ and $V$ are simple Hermitian operators which are defined as follows
\be\label{eq:VW}
W=\O_\phi+c_1 \O_\xi\ , \qquad V=\O_\phi+c_2 \O_\xi\ ,
\ee 
where $c_1$ and $c_2$ are arbitrary real numbers. In the Regge limit (\ref{regge2}), these type of correlators obey some nice properties in any unitary CFT. For example, any Lorentzian correlators, such as $G$, where operators are inserted symmetrically in the Rindler wedges can be interpreted as thermal correlators. More precisely, the Minkowski vacuum can be interpreted as the thermofield double, entangling the right Rindler wedge with the left Rindler wedge. In this ``thermal" state of temperature $T$, a standard measure  for chaos is the out-of-time-order correlator (OTOC)  \cite{Maldacena:2015waa}
\be
F(t_R)=\tr\[ yVyW(t_R)yVyW(t_R)\]
\ee
where $t_R$ in this case is the Rindler time and 
\be
y^4=\frac{e^{-\beta H}}{\tr\[e^{-\beta H}\]} \ , \qquad \beta=\frac{1}{T}\ .
\ee
The Minkowski correlator \eqref{eq:correlator} now can  be viewed as a thermal correlator on Rindler space
\be\label{eq:otoc}
G=\frac{F\(t_R-\frac{i\beta}{4}\)}{F_d}\ , \qquad F_d=\tr\[y^2 V y^2 V\]\tr\[y^2 W(t_R) y^2 W(t_R)\]\ ,
\ee
where,  $e^{2\pi t_R/\beta}=\rho/\sqrt{\rho \bar{\rho}}$. The correlator $G$, in the Regge limit, behaves in the following way 
\be
G=1+\delta G\ ,
\ee
where the growth of $\delta G\sim \rho/\sqrt{\rho \bar{\rho}} $ can now be thought of as the Lyapunov growth of a thermal quantum system. Moreover, $\Delta_f$ has now has a natural interpretation as the scrambling time $t_* = \beta \log(\Delta_f)$.

The chaos bound of \cite{Maldacena:2015waa} imposes  rigorous constraints on $\delta G$ in the Regge limit (\ref{regge2}):  (i) $\delta G$ must not grow faster than $\r$,  (ii) when  $\delta G$ grows as $\r$
\be\label{chaos}
\mbox{Im}\ \delta G\ge  0 \qquad \text{for} \qquad 0<\r\br<1\ .
\ee
The chaos bound can be interpreted as a causality constraint \cite{Afkhami-Jeddi:2016ntf,Afkhami-Jeddi:2017rmx,Afkhami-Jeddi:2018own} or as a unitarity constraint in certain scenarios \cite{Kulaxizi:2017ixa, Costa:2017twz}. The positivity condition \eqref{chaos} applies to effective CFTs as well since it follows from a CFT sum-rule.\footnote{This CFT sum-rule plays a crucial role in constructing $C$-functions from OTOC. We will discuss this in section \ref{sec:C}.}

We are now in a position to compute $\delta G$ by utilizing our results from section \ref{sec:regge}. Specifically, we obtain
\begin{align}
 G= &\frac{(16\r\br)^3}{c_0^2\(1+c_1^2\)\(1+c_2^2\)}\(G_{\phi\phi\phi\phi}+c_1^2 c_2^2 G_{\xi\xi\xi\xi}+\(c_1^2+c_2^2\)G_{\phi\phi\xi\xi}+4c_1 c_2 G_{\phi\xi\phi\xi}\)
\end{align}
implying $\delta G \sim i \r \sim i e^{2\pi t_R/\beta}$. The chaos bound (\ref{chaos}) now imposes
\be\label{final_bound}
8\Delta a + 4c_1^2c_2^2 B+\(c_1^2+c_2^2\)b+2c_1c_2\(b-8\Delta a\)\ge 0
\ee
for all $c_1,c_2 \in \mathbb{R}$. First, the above inequality immediately implies the 4d $a$-theorem
\be
\Delta a= a_{UV}-a_{IR}\ge 0\ ,
\ee
where the equality holds only when the dilaton is exactly free representing the case in which there is no RG flow.  Moreover, the inequality (\ref{final_bound}) imposes constraints on $B$ and $b$ as well
\begin{align}\label{eq:comment}
 B\ge 0\ , \qquad b\ge 0 \ , \qquad b\ge 4\Delta a-\sqrt{8B\Delta a}\ .
\end{align}
Note that $b=0$ and/or $B=0$ necessarily require $\Delta a=0$ (no RG flow). The excluded region in the $B-b$ plane is shown in figure \ref{intro:chaos}. The last inequality which follows from the {\it interference effect} can be interpreted as an upper bound on $\Delta a$. 

Clearly, $\delta G$ is a monotonically increasing function of Rindler time $t_R$. This fact, as we will explain in section \ref{sec:C}, is closely related to the existence of multiple $C$-functions that decrease monotonically under RG flows in 4d.  

As mentioned in the introduction, the bounds (\ref{eq:comment}) can also be obtained directly in flat space following \cite{Adams:2006sv} with some assumption about the asymptotic behavior of four-point scattering amplitudes. The last bound of (\ref{eq:comment}) is more subtle and may require additional assumptions (see \cite{Remmen:2019cyz,Bellazzini:2020cot,Tolley:2020gtv,Trott:2020ebl} for similar bounds).

\subsection{Bootstrap Corner}
When we look closely, there is a {\it kink} in the exclusion plot \ref{intro:chaos}. The kink is located at\footnote{As mentioned before, $b=0$ is ruled out. By equation (\ref{eq:comment2}), we mean that $\frac{b}{\Delta a}\rightarrow 0$ is parametrically suppressed.} 
\be\label{eq:comment2}
B=2\Delta a\ , \qquad b=0
\ee
which corresponds to the effective action
\begin{align}\label{corner}
S_{\text{eff}}[\phi,\xi]=&\int d^4x \(-\frac{1}{2}\(\p \phi\)^2-\frac{1}{2}\(\p \xi\)^2+\frac{2\Delta a}{f^4}\(\(\p \phi\)^2-\(\p \xi\)^2\)^2\)\ .
\end{align}
This type of corner, often seen in the conformal bootstrap, is associated with interesting theories. However,  we are not aware of any RG flows that are described by the effective action \eqref{corner}.

\section{Supersymmetric Flows}\label{sec:susy}
\begin{figure}
\centering
\includegraphics[scale=0.3]{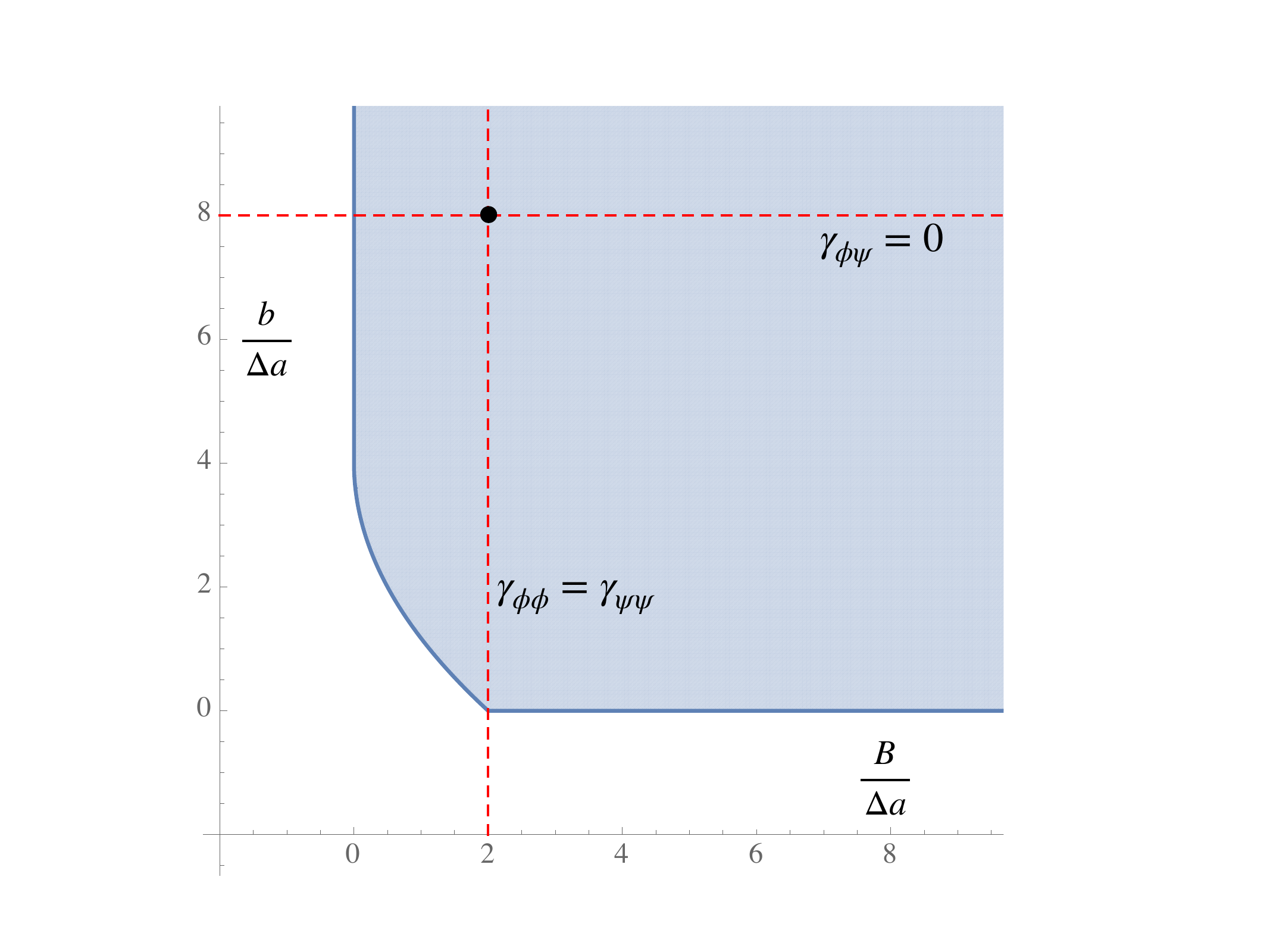}
\caption{ \label{susy} \small 4d RG flows connecting two $\N=1$ SCFTs are represented by the black dot inside the allowed region in $B$-$b$ plane. Two dashed red lines correspond to the $\N=1$ supersymmetric Ward identities which can be equivalently  stated as $\gamma_{\phi\phi}=\gamma_{\xi\xi}$ and $\gamma_{\phi\xi}=0$.}
\end{figure}

A simple example of an RG flow with broken global symmetry comes naturally from supersymmetry. Consider 4d $\N=1$ SCFTs in which conformal symmetry is broken by an operator that preserves the $\N=1$ supersymmetry. This breaks the $U(1)$ R-symmetry as well since the stress tensor is in the same supermultiplet as the R-current. As a result the theory flows to another SCFT in the deep IR. In this scenario, the NG fields $\tau$ and $\beta$ are part of a chiral superfield $\Phi=\tau+i\beta+\cdots$. The resulting low energy effective action for the bosonic part is given by \cite{Schwimmer:2010za,Bobev:2013vta}
\begin{align}\label{action_susy}
S_{\text{eff}}[\phi,\xi]=\frac{1}{2}\int d^4x \(-\(\p \phi\)^2-\(\p \xi\)^2+\frac{\Delta a}{f^4}\(\phi^2 \Box^2 \phi^2+\xi^2 \Box^2 \xi^2-2\xi^2 \Box^2 \phi^2+4 \phi \xi \Box^2 \phi \xi\)\)
\end{align}
implying 
\be\label{eq:susy}
B=2\Delta a\ , \qquad b=8\Delta a\ .
\ee
These relations can be thought of as the $\N=1$ supersymmetric Ward identities \cite{Bobev:2013vta}. The relations (\ref{anomalous}) allow us to reinterpret these Ward identities as simple relations among various anomalous dimensions in the dual CFT$_3$
\be\label{bound:susy}
\gamma_{\phi\phi}=\gamma_{\xi\xi}\ , \qquad \gamma_{\phi\xi}=0
\ee 
as shown in figure \ref{susy}.

In section \ref{sec:C}, we will argue that there are infinitely many distinct $C$-functions that decrease monotonically from $a_{\rm UV}$ to $a_{\rm IR}$ under RG flows. In particular, for $\N=1$ supersymmetric flows in 4d, by using the Ward identities of \cite{Bobev:2013vta}, we  define the following set of independent $C$-functions (also known as $a$-functions) in the dual CFT$_3$ description: 
\begin{align}\label{def:C_susy}
C(\mu)=a_{\rm IR}+\frac{\tilde{\Delta}_f^4 \eta^{\frac{7}{2}}}{f_{3333}\(-\frac{1}{2}\log (\eta)\)}\lim_{\frac{1}{\Delta_f^4}\ll x\ll 1} \int^{x}_{\Delta_f^{-4\mu/f}x} d\sigma\ \mbox{Re}\(\frac{(r_1^2+r_2^2)c_0^2}{(16\eta)^3}-r_1^2G_{\phi\phi\phi\phi}-r_2^2G_{\phi\phi\xi\xi}\)
\end{align}
for all $r_1,r_2 \in \mathbb{R}$, where $\eta$ and $\sigma$ are defined in equation (\ref{def:sigma}).\footnote{Note that $c_1, c_2$ and $r_1,r_2$ are related in the following way
\be\nonumber
r_1=(1+c_1 c_2)\ , \qquad r_2=c_1-c_2\ .
\ee
Moreover, as we found before $c_0=\frac{12}{\pi^2}$.} In the above equation, we have exploited the positivity of the integrand which  follows from Rindler positivity \cite{Hartman:2016lgu}. Moreover, $\tilde{\Delta}_f$ is given by $f R_{\rm AdS}$ times some positive numerical factor (which is independent of $\eta$, $r_1$, and $r_2$). The numerical factor can always be chosen such that
\be
C\(\frac{\mu}{f}\rightarrow 0\)=a_{\rm IR} \ , \qquad C\(\frac{\mu}{f}\rightarrow \infty\)=a_{\rm UV}\ .
\ee
This will be discussed in more detail in section \ref{sec:C}, where the integral in the right hand side of (\ref{def:C_susy}) will be written as an integral over an OTOC. Of course, a similar set of $C$-functions can also be constructed from combinations of flat space amplitudes $\mathcal{A}(\phi\phi\phi\phi)$ and $\mathcal{A}(\phi\xi\phi\xi)$ by extending the procedure presented in \cite{Komargodski:2011vj}. 

The presence of multiple $a$-functions for $\N=1$ flows has a long and interesting history. For example, it was a source of much interest right after $a$-maximization was proposed. However, it is not clear if there is any relation between the $C$-functions of (\ref{def:C_susy}) and $a$-maximization \cite{Intriligator:2003jj,Kutasov:2003iy,Kutasov:2003ux}. In fact, it is also not obvious whether different $C$-functions of (\ref{def:C_susy}) are truly distinct. It is possible that these $C$-functions just represent different RG schemes.\footnote{We thank J. Heckman for pointing it out.}

\section{Generalization to Non-Abelian Global Symmetries}\label{sec:general}

There are a few subtleties  associated with generalizing the preceding discussion to the breaking of  non-abelian global symmetries. Now we start with a $\CFTUV$ in (3+1)-dimensions with a  global symmetry group $G$, where $G$ is any compact Lie group. For simplicity, we assume that $G$ is simple. However, as we will explain later, our result is applicable even when $G$ is a direct product of finite number of simple Lie groups.

Similar to the abelian case, we again couple the UV theory to a background metric $g_{\mu\nu}(x)$ and a background gauge field $A_\mu(x)$. In the present case, it is important that the gauge field $A_\mu(x)$ is introduced in such a way that it makes the global symmetry of the UV theory local. In general, the global symmetry $G$ can have 't Hooft anomaly. In that case, the above gauging seems to be problematic. However, we can always introduce a set of massless spectating fields which only couple to the gauge field $A_\mu$ (but not $\CFTUV$) in such a way that the $G$-anomaly is cancelled. This standard trick enables us to make the global symmetry of $\CFTUV$ local by coupling it to a gauge field $A_\mu$. Of course, at the linearized level, $A_\mu$ is coupled to $\CFTUV$ through the spin-1 conserved current associated with the global symmetry $G$.

Let us now imagine that the conformal symmetry and the global symmetry of  $\CFTUV$ are spontaneously broken by turning on VEVs for some operator
\be
G \times \mathfrak{so}(4,2)\rightarrow  \mathfrak{iso}(3,1)\ .
\ee
This starts an RG flow that ends at $\CFTIR$. We will only consider RG flows in which $\CFTIR$ is invariant under the action of the group $G$. This means either the ``individual fields" of $\CFTIR$ transform trivially under the group $G$ or more generally $\CFTIR$ also has the UV symmetry $G$ (which can be anomalous). Equivalently, the gauge field $A_\mu$ at the linearized level, couples only to some spin-1 conserved current of $\CFTIR$ (if at all).\footnote{Note that the massless spectator fields that were introduced to cancel the UV $G$-anomaly will survive even at the IR. The full IR theory, including the spectators, must be free from $G$-anomaly. Hence, we can simply incorporate the effects of these spectator fields by implementing the 't Hooft anomaly matching condition. } Of course, $\CFTIR$ can have other global symmetries that do not embed at all in the UV theory. 
 
 These class of RG flows are also described in terms of an effective action of the NG bosons of broken symmetries. The same effective action also describes RG flows where the conformal symmetry and the global symmetry of the $\CFTUV$ are broken explicitly. In that scenario, as discussed before, the NG bosons should be interpreted as compensator fields. 

\subsection{Effective Action}
One advantage of coupling the theory to background fields is that the standard coset construction for spontaneous symmetry breaking emerges naturally from it. Moreover, it is also more convenient to track all anomalies when we couple the theory to a background metric $g_{\mu\nu}(x)$ and a background gauge field $A_\mu(x)$. The background gauge field $A_\mu(x)$ can be decomposed as follows
\be
A_\mu(x)= A_\mu^i(x) T^i\ ,
\ee
where $T^i$ with $ i\in\{1,2,\cdots,\mbox{dim}\ G\}$ are Hermitian generators of $G$ in the fundamental representation satisfying
\be\label{eq:lie}
[T^i,T^j]=i f^{ijk}T^k
\ee
and $\mbox{Tr}\ T^i T^j \propto \delta^{ij}$.

The broken global symmetry generates massless NG bosons $\beta_i$, $ i\in\{1,\cdots,\mbox{dim}\ G\}$, which accompany the dilaton $\tau$. The low energy effective action $S_{\rm eff}[\tau,\beta_i]$ can be derived by studying the variation of the action under diff$\times$Weyl transformations and gauge transformations. Weyl transformations act in the usual way (\ref{weyl}). On the other hand, the gauge field transforms under the gauge transformation as
 \begin{align}\label{eq:gauge}
 A_\mu(x) \rightarrow \Omega(x)  A_\mu(x) \Omega^{-1}(x)+i \Omega(x) \p_\mu \Omega^{-1}(x)\ ,
 \end{align}
 where $\Omega(x)=e^{i \alpha_i(x) T^i} \in G$. Under the same gauge transformation, NG fields $\beta(x)\equiv \beta_i(x) T^i$ transform as
 \be
g(x)\equiv e^{i\beta(x)}\rightarrow \Omega(x) e^{i\beta (x)}\ .
 \ee 
 The infinitesimal gauge transformation takes the familiar form 
 \be\label{gauge2}
A_\mu(x) \rightarrow A_\mu(x) +\p_\mu \alpha(x)-i [A_\mu,\alpha]\equiv A_\mu(x) +{\mathcal D}_\mu \alpha(x)\ , \qquad \beta(x)\rightarrow\beta(x)+ \alpha(x)\ ,
\ee
where $\alpha(x)=\alpha^i (x)T^i$. 

We now can simply repeat the argument of the $U(1)$ case. The anomaly matching arguments of \cite{Schwimmer:2010za,Komargodski:2011vj} applies here as well implying that the changes in anomalies  in the flow from  $\CFTUV$ to $\CFTIR$ must be compensated by the NG bosons. This leads to the variation of  the effective action under  an infinitesimal  Weyl transformation 
\be\label{variation3}
\delta_\sigma S_{\rm eff}[g_{\mu\nu},A_\mu;\tau,\beta_i]=\int d^4x \sqrt{-g} \sigma(x)\left(-\Delta a E_4+  \Delta c W^2+\Delta \kappa_G \mbox{Tr} F^2\right)
\ee
where $\kappa_G$ is the trace anomaly associated with the background gauge field $A_\mu(x)$. Similarly, the variation of the effective action under an infinitesimal gauge transformation (\ref{gauge2}) must have the following form
\be\label{variation4}
\delta_\alpha S_{\rm eff}[g_{\mu\nu},A_\mu;\tau,\beta_i]=\int d^4x \sqrt{-g}\left(\Delta c_A  d_{ijk}\alpha^i F^j_{\mu\nu}\tilde{F}_k^{\mu\nu}+\Delta c_g \mbox{Tr}( \alpha) R_{\mu\nu\alpha\beta}\tilde{R}^{\mu\nu\alpha\beta}\right)\ ,
\ee
where, $c_A$ and $c_g$ are anomaly coefficients and $d_{ijk}=\mbox{Tr} \(\{T_i,T_j\}T_k\)$. Note that the second term in (\ref{variation4}) vanishes since $\mbox{Tr}\ T_i=0$. Hence, there is no mixed gauge-gravitational anomaly when the symmetry group $G$ is simply laced. Nevertheless, we kept both terms since later we will generalize to symmetry groups that may contain $U(1)$ factors. This does not cost us anything because  in the flat space limit with no background gauge field, the gauge anomalies \eqref{variation4} do not contribute to the low energy effective action of NG bosons.

The rest of the argument is exactly the same as before implying that in the flat space limit with no background gauge field $S_{\text{eff}}[\tau,\beta_i]$ still has the same simple form
\be\label{general}
S_{\text{eff}}[\tau,\beta_i]=\(S_{\rm WZ}+S_{\rm inv}\)_{g_{\mu\nu}=\eta_{\mu\nu},A_\mu=0}\ ,
\ee
where, the Wess-Zumino part of the action $S_{\rm WZ}$ produces both the conformal anomaly \eqref{variation3}, as well as the anomaly for the global symmetry $G$ \eqref{variation4}. Similar to the abelian case, $S_{\rm WZ}$ in the flat space limit with no background gauge field is uniquely fixed by $\Delta a$
\be
S_{\rm WZ}|_{g_{\mu\nu}=\eta_{\mu\nu},A_\mu=0}=2 \Delta a\int d^4x \sqrt{-g}\(\nabla \tau\)^2\(2\Box \tau-\(\nabla \tau\)^2\)\ .
\ee
Furthermore, $S_{\rm inv}$ can be constructed from the Weyl invariant combination $\hat{g}_{\mu\nu}=e^{-2\tau}g_{\mu\nu}$ and the gauge covariant combination $\hat{A}_\mu=A_\mu -\omega_\mu$, where 
\be\label{def:MC}
\omega_\mu(x) \equiv i g(x) \p_\mu g^{-1}(x) \equiv \omega_\mu^i(x) T_i
\ee 
with $g(x)\equiv e^{i\beta(x)}$. Under the gauge transformation we find $\hat{A}_\mu\rightarrow \Omega(x)  \hat{A}_\mu \Omega^{-1}(x) $. Note that $\omega_\mu(x)$ is precisely the Maurer-Cartan form  which  plays a central role in the coset construction. This allows us to define a coset covariant derivative 
\be\label{eq:cov}
\omega_\mu(x) = D_\mu \beta(x)\ .
 \ee
Therefore, up to four-field level, we can write 
\begin{align}\label{inv}
S_{\rm inv}=\int d^4x \sqrt{-\hat{g}}\left(-\frac{f^2}{2}\( \frac{\hat{R}}{6}+2 \gamma_0^2 \hat{g}^{\mu\nu}\Tr\(\hat{A}_\mu\hat{A}_\nu\)\) +\sum_{I=1}^{10} \gamma_I \tilde{W}_I+\O(\p^6)\right)\ ,
\end{align}
where,  $ \tilde{W}_I$ are all four-derivative invariants which are given in appendix \ref{app:NA}. Similar to the previous sections, the mass scale $f$ represents the symmetry breaking scale. Whereas, $\gamma$-coefficients are real, dimensionless, and theory dependent.

After using the equations of motion and  taking the flat space limit with no background gauge field, we obtain the low-energy effective action (\ref{app:final}) for the {\it physical dilaton} $\phi$ and {\it physical axions} $\xi_i$ (for details see appendix \ref{app:NA}). The effective action at the four-derivative and four-field level has the form (\ref{intro:action})
\be\label{eq:eff_act}
S_{\text{eff}}[\phi,\xi_i]=S_{\rm conformal}[\phi; \Delta a]+S_{\rm global}[\xi_i; B_{ijkl}]+S_{\rm mixed}[\phi,\xi_i;\Delta a, b]\ .
\ee
As noted in the introduction, the effective action of the dilaton $S_{\rm conformal}[\phi; \Delta a]$ remains unaffected by breaking of the global symmetry $G$
\be\label{eff:conformal}
S_{\rm conformal}[\phi; \Delta a]=\int d^4x  \(-\frac{1}{2}\(\p \phi\)^2+\frac{\Delta a}{2f^4}\phi^2 \Box^2 \phi^2\)+ \O\(\p^6;\phi^6\)
\ee
which agrees with the dilaton effective action of \cite{Komargodski:2011vj}. Similarly, $S_{\rm global}[\xi_i; B_{ijkl}]$ is the axionic part of the effective action\footnote{Let us note that $S_{\rm global}[\xi_i; B_{ijkl}]$ can alway be written in the form (\ref{eq:global}). However, this requires scaling the generators such that $\Tr\(T_i T_j\)=\frac{1}{2\gamma_0^2} \delta_{ij}$, where $\gamma_0>0$ is theory dependent. The structure constants $f_{ijk}$ are also defined in this convention.} 
\begin{align}\label{eq:global}
S_{\rm global}[\xi_i; B_{ijkl}]=& -\frac{1}{2}\int d^4x \left( \p \xi_i\cdot \p \xi_i-\frac{1}{6f^2 }f_{ijk}f_{ij'k'} \xi_j \xi_{j'}\Box \(\xi_k  \xi_{k'}\)+\frac{1}{4f^2 }\sum_{i\neq j}\xi_i^2\Box \xi_j^2\)\nonumber\\
&+\frac{1}{4f^4}\int d^4x   B_{ijkl} \xi_i  \xi_j \Box^2 \( \xi_k  \xi_l\) +\O\(\p^6;\xi^6\)\ ,
\end{align}
where, $ B_{ijkl}$ is fixed by symmetry 
\be\label{def:B}
B_{ijkl}=B_1 \delta_{ij}\delta_{kl}+B_2 \(\delta_{ik}\delta_{jl}+\delta_{il}\delta_{jk}\)+B_3\( f_{i'ik}f_{i'jl}+f_{i'il}f_{i'jk}\)+B_4T_{ijkl}
\ee
up to arbitrary dimensionless coefficients $B_1,B_2,B_3,$ and $B_4$.  Note that we have defined $T_{ijkl}=\Tr\(\{T_i, T_j\}\{T_k,T_l\}\)$ and $f_{ijk}=-2i \Tr\([T^i,T^j]T^k\)$. Of course, for a specific $G$ all $B$-coefficients may not be independent. For example, for $G=SU(2)$ it is sufficient to set $B_3=B_4=0$.  

In contrast to the abelian case, the axionic part of the effective action (\ref{eq:global}) also contains two-derivative four-field interactions.  This should not be surprising since spontaneous breaking of a non-abelian continuous global symmetry can generate two-derivative four-field interactions which follow directly from the Maurer-Cartan form (\ref{def:MC}). In fact, this type of two-derivative interactions are already present in more familiar chiral Lagrangians in particle physics which lead to radiative corrections. However, there is a crucial difference. The term $\xi_i^2\Box \xi_j^2$ in the action (\ref{eq:global}) appears only when the breaking of global symmetry is accompanied by a breaking of conformal symmetry. In other words, taking the physical dilaton $\phi=0$ in (\ref{eq:eff_act}) does not reproduce the low energy effective action associated with spontaneous breaking of only the global symmetry $G$. On the other hand, the limit $\tau=0$ is actually smooth reproducing the correct effective action for the broken global symmetry.

Finally, the mixed part of the action $S_{\rm mixed}[\phi,\xi_i;\Delta a,b]$ represents interactions between $\phi$ and $\xi_i$
\begin{align}
S_{\rm mixed}[\phi,\xi_i;\Delta a,b]=\frac{1 }{4f^4} \int d^4x \(b  \phi \xi_i \Box^2 \( \phi  \xi_i\)-4\Delta a \xi^2 \Box^2 \phi^2  \) + \O\(\p^6;\phi^2\xi^4;\phi^4\xi^2 \)
\end{align}
where, $b$ is a dimensionless coupling constant. Note that the second term of $S_{\rm mixed}$ is universal. In general, $b$ and $B$-coefficients depend on $\CFTUV$, $\CFTIR$, and deformations (or VEVs) that trigger the RG flow. As the abelian case leads us to expect, these coefficients must also satisfy various positivity conditions for unitary RG flows which we will derive next. However, presence of the  two-derivative four-field interactions in (\ref{eq:global}) makes these bounds more subtle.

\subsection{Bounds from Chaos}
Similar to the $U(1)$ case, we again consider the dual CFT$_3$ description of RG flows characterized by (\ref{eq:eff_act}). The dual CFT$_3$ now contains $N+1$ scalar primary operators of dimensions $\Delta=3$.  The operator dual to the physical dilaton is denote by $\O_\phi$. Similarly,  operators $\O_i$ for $i=1,\cdots, N$ are dual to $\xi_i$.  In this dual description, consider the Regge correlator (\ref{eq:correlator}) in the kinematics (\ref{points}), where operators $W$ and $V$ now are defined as follows
\be\label{def:operators}
W=\O_\phi+\sum_i c_i \O_i\ , \qquad V=\O_\phi+\sum_i \c_i \O_i\ ,
\ee 
where $c_i$ and $\c_i$ are arbitrary real numbers. The chaos bound (\ref{chaos}) now imposes
\begin{align}\label{eq:bound}
8\Delta a + b\sum_i \(c_i^2+\c_i^2\) +2\(b-8\Delta a\)\sum_i c_i\c_i+4\sum_{i,j,k,l} B_{ijkl}^{\rm eff}c_i\c_j c_k\c_l \ge 0
\end{align}
for all $c, \c \in \mathbb{R}^N$. Unlike the abelian case, now there can be loop contributions 
\be\label{eq:Beff}
B_{ijkl}^{\rm eff}= B_{ijkl}+B_{ijkl}^{\rm 1-loop}
\ee
where $B_{ijkl}^{\rm 1-loop}$ represents 1-loop contributions to the Regge correlator (\ref{eq:correlator}) from the two-derivative four-field interactions in (\ref{eq:global}).

\subsubsection{Linear Constraints}
First, note that the consistency condition (\ref{eq:bound}) imposes 
\begin{align}\label{eq:b}
\Delta a>0 \ , \qquad b> 0\ , 
\end{align}
implying that the broken global symmetry does not affect the proof of the $a$-theorem.  Similarly, we also find that $B_{ijkl}^{\rm eff}$ is {\it strongly elliptic}. In other words, it has a positive definite bi-quadratic form 
\be\label{eq:elliptic}
Bc^2\c^2\equiv \sum_{i,j,k,l}B_{ijkl}^{\rm eff}c_i\c_jc_k\c_l > 0
\ee
for all $c,\c\in \mathbb{R}^N$ over unit spheres $\sum_i c_i c_i=\sum_i \c_i\c_i=1$. It should be noted that in general the constraint (\ref{eq:elliptic}) is not very interesting since it is automatically satisfied because of the 1-loop contributions from the two-derivative four-field interactions in (\ref{eq:global}). From the effective field theory perspective the two-derivative four-field interactions in (\ref{eq:global}) induce log runnings of $B_1$, $B_2$, $B_3$, and $B_4$ which dominate at low energies. These log runnings ensure that the constraint (\ref{eq:elliptic}) is  trivially satisfied. This is very similar to the constraints on the $SU(2)$ chiral Lagrangian, as discussed in  \cite{Pham:1985cr,Adams:2006sv}.

On the other hand, if the two-derivative four-field interactions in (\ref{eq:global}) are parametrically suppressed because of weak coupling $|B_{ijkl}|\gg |B_{ijkl}^{\rm 1-loop}|$, the constraint (\ref{eq:elliptic}) becomes non-trivial.\footnote{In the effective action (\ref{eq:global}), this weak coupling suppression can be equivalently stated as $|B_1|,\cdots,|B_4|\gg 1$ but not too large so that they do not affect the perturbative expansion in $\frac{1}{f}$.} In this case $B_{ijkl}^{\rm eff} \approx B_{ijkl}$ and hence the positivity constraint (\ref{eq:elliptic}) leads to interesting bounds on $B$-coefficients. Clearly, when loop contributions are suppressed, the positivity condition (\ref{eq:elliptic}) is non-trivial and holds whenever any global symmetry is spontaneously broken with or without conformal symmetry breaking.

\subsubsection{Nonlinear Constraints}
When loop contributions are suppressed $|B_{ijkl}|\gg |B_{ijkl}^{\rm 1-loop}|$, there are stronger conditions that $\Delta a$, $b$, and  $B_{ijkl}$ must also satisfy. In particular, the relation (\ref{eq:bound}) imposes that 
\begin{align}\label{def:A}
A[c,\c]\equiv \Delta a\(1-2c\cdot \c\)+\frac{1}{8} b \(c +\c\)^2+\frac{1}{2} (B_1+B_2)\(c\cdot \c\)^2 +\frac{1}{2} B_2 c^2 \c^2 \nonumber\\
+\frac{1}{2}\(B_3 f_{i'ij}f_{i'kl}+B_4 T_{ijkl}\)c_i \c_j c_k \c_l \ge 0
\end{align}
for all $c,\c\in \mathbb{R}^N$. Clearly, for $c\cdot \c=\sum_i c_i\c_i>0$ the above condition is stronger than the previous positivity conditions when loop contributions are absent. Of course, these infinite set of inequalities are not all independent. It is always possible to reduce $A[c,\c]\ge 0$ into a finite number of constraints on $\Delta a$, $b$ and  $B_I$.

\subsection{Example: $SU(2)$}
Let us provide an example to persuade the reader that for specific theories  the above constraints simplify greatly. We consider a scenario in which  $\CFTUV$ has a global $SU(2)$ symmetry. The conformal symmetry and the global symmetry of $\CFTUV$ are explicitly or spontaneously broken 
\be
SU(2) \times \mathfrak{so}(4,2)\rightarrow  \mathfrak{iso}(3,1)
\ee
which starts an RG flow that ends at $\CFTIR$. This happens naturally when the $\CFTUV$ has $\N=2$ supersymmetry since   $\N=2$ SCFTs have $U(1)\times SU(2)$ R-symmetry.

The $SU(2)$ symmetry implies 
\be
B_{ijkl}=B_1 \delta_{ij}\delta_{kl}+B_2 \(\delta_{ik}\delta_{jl}+\delta_{il}\delta_{jk}\)\ .
\ee
Of course, the constraints (\ref{eq:b}) remain unaffected. On the other hand, the condition that $B_{ijkl}$ is strongly elliptic imposes  
\be\label{su2:B}
B_1+B_2> 0\ , \qquad B_2> 0\ .
\ee
As discussed before, the positivity conditions (\ref{su2:B}) are non-trivial only when 1-loop contributions are negligible. Furthermore, when 1-loop contributions are suppressed, $A[c,\c]\ge 0$ reduces to a stronger nonlinear constraint 
\begin{align}
b\ge 4\Delta a -\sqrt{8\Delta a (B_1+2B_2)} 
\end{align}
which can be viewed as an upper bound on $\Delta a$. Clearly, we again obtain an exclusion plot which is exactly the same as   figure \ref{intro:chaos} with the substitution $B\rightarrow B_1+2B_2$. Furthermore, this identification suggests that RG flows between $\N=2$ SCFTs in which conformal symmetry and $SU(2)_R$ symmetry are broken by an operator that preserves the $\N=2$ supersymmetry, are described by $b=8\Delta a$ and $B_1+2B_2=2\Delta a$. It would be nice to verify this expectation directly from supersymmetric Ward identities.

\subsection{A More General Scenario}
So far, we have assumed that $G$ is a simple Lie group. However, the  form of the effective action (\ref{general}) implies that the same analysis holds for a more general scenario. In particular, the preceding argument applies even when $G$ is a direct product of finite number of simple Lie groups and $U(1)$
\be
G=\prod_a G_a\ .
\ee
Now there can be mixed anomalies associated with various $G_a$. However, contributions from these additional anomalies also vanish when we take the flat space limit with no background fields. In general, the RG flow between $\CFTUV$ and $\CFTIR$ can preserve some subgroup $H$ of $G$. In such a scenario $\CFTIR$ has the global symmetry $H$. In fact, it is possible that deep in the IR  some of the broken UV symmetries get restored and hence the $\CFTIR$ can have a bigger symmetry group $ H'\ge H$. This situation, for example, arises naturally for supersymmetric flows. We can include this possibility as well since the low energy effective action of the NG bosons depends only on $H$. In general, these RG flows are also described in terms of a low energy effective action of a dilaton $\phi$ and $N$ axions $\xi_i$ where 
\be\label{defN}
N=\mbox{dim}\ G-\mbox{dim}\ H=\sum_a \mbox{dim}\ G_a-\mbox{dim}\ H\ .
\ee
The effective action still has the form (\ref{eq:eff_act}) where $S_{\rm conformal}$ remains unaffected (\ref{eff:conformal}). A straightforward generalization leads to
\begin{align}\label{Gaction}
S_{\text{eff}}[\phi,\xi_i]&= \int d^4x \(-\frac{1}{2}\left(\(\p \phi\)^2+ \sum_{i=1}^N \(\p \xi_i\)^2\)+\frac{\Delta a}{2f^4} \(\phi^2\Box^2\phi^2  - 2 \sum_{i=1}^N  \xi_i^2 \Box^2 \phi^2\)+\mathcal{L}_{(2)}[\xi]\)\nonumber \\
&+\frac{1}{4f^4}\int d^4x \( \sum_{i=1}^N b_{i}\phi\xi_i\Box^2\phi\xi_i+ \sum_{i,j,k,l=1}^N B_{ijkl}\xi_i\xi_j\Box^2\xi_k\xi_l\) \ ,
\end{align}
where $N$ is given by  (\ref{defN}). Again notice that NG modes $\xi_i$ of broken global symmetries do not interfere with the proof of the 4d $a$-theorem. In this general case, the low energy effective action of NG bosons is completely fixed by symmetry up to dimensionless coupling coefficients $\{b_i, B_{ijkl}\}$.\footnote{Note that just from symmetry argument we get a real symmetric matrix $b_{ij}$ and a 4-tensor $B_{ijkl}$. We can always perform a field redefinition to diagonalize $b_{ij} \Rightarrow b_i \delta_{ij}$.} Just like before, coupling coefficients  $B\equiv \{B_{ijkl}\}$ is a strongly paired symmetric 4-tensor which has the symmetries of the $N$-dimensional {\it elasticity tensor}
\be
B_{ijkl}=B_{ji kl}=B_{ijlk}=B_{klij}\ .
\ee
However, now it may not have the form (\ref{def:B}) in general. Together, $\{b_i, B_{ijkl}\}$ contains $\frac{1}{8} N \left(N^3+2 N^2+3 N+10\right)$ independent coefficients. Although, in a specific theory some of these coefficients can be related to each other and/or $\Delta a$. 

In the above effective action (\ref{Gaction}), the axionic part also contains two-derivative four-field interactions $\mathcal{L}_{(2)}[\xi]$. This can be obtained directly from our earlier analysis 
\be\label{def:L2}
\mathcal{L}_{(2)}[\xi]=\frac{1}{12f^2 }f_{ijk}f_{ij'k'} \xi_j \xi_{j'}\Box \(\xi_k  \xi_{k'}\)-\frac{1}{8f^2 }\sum_{i\neq j}\xi_i^2\Box \xi_j^2\ ,
\ee
where, axions $\xi_i$ with $i\in \{1,2,\cdots,N\}$ belong in some large reducible representation with structure constants $f_{ijk}$. Clearly, the first term vanishes for NG bosons associated with broken $U(1)$s.

The chaos bound now leads to a similar positive function in the space of $\{\Delta a, b_i,B_{ijkl}\}$
\be\label{def:A1}
A[c,\c]=\Delta a\(1-2c\cdot \c\)+\frac{1}{8} \sum_{i=1}^N b_i \(c_i+\c_i\)^2+\frac{1}{2}\sum_{i,j,k,l}B_{ijkl}^{\rm eff}c_i\c_jc_k\c_l\ge 0
\ee
for all $c,\c\in \mathbb{R}^N$, where $B_{ijkl}^{\rm eff}$ is defined as before (\ref{eq:Beff}). This positivity condition leads to $\Delta a> 0$, $b_i> 0$, and $B_{ijkl}^{\rm eff}$ is strongly elliptic (\ref{eq:elliptic}). Some of these constraints involving $B_{ijkl}$ can be trivially satisfied because of the 1-loop contributions from the two-derivative four-field interactions (\ref{def:L2}). 

This is the most we can say about the effective action (\ref{eq:eff_act}) without requiring to know anything about the details of the flow. For specific theories, some components of $\{\Delta a, b_i,B_{ijkl}\}$ can actually be related. In that case, the above constraints simplify greatly. In general, the condition (\ref{def:A1}) can be alternatively and equivalently stated as an eigenvalue problem \cite{Qi1}
\be\label{eq:eigen}
\sum_{j,k,l} A_{ijkl}\c_jc_k\c_l =\lambda c_i\ , \qquad \sum_{i,j,k} A_{ijkl}c_i\c_jc_k =\lambda \c_l\ , \qquad \sum_i c_i c_i=\sum_i \c_i\c_i=1
\ee
where $c$ and $\c$ are left and right eigenvectors of $A$ with eigenvalue $\lambda \in \mathbb{R}$. Now the condition (\ref{eq:elliptic}) implies that all eigenvalues of (\ref{eq:eigen}) must be positive. In spirit, this is analogous to the matrix eigenvalue problem, however, in general the eigenvalue problem (\ref{eq:eigen}) for $N>3$ is difficult to solve. In fact, it is known that the optimization problem (\ref{eq:elliptic}) is NP-hard \cite{Qi2}.

\section{Sum-Rules, OTOC and C-Functions}\label{sec:C}
It was argued in \cite{Kundu:2019zsl} that some properties of RG flows are more transparent in the dual CFT description in one less dimension. For example we can write a rigorous CFT$_3$ sum-rule for $A[c,\c]$, as defined in \eqref{def:A1} (with (\ref{final_bound}) and (\ref{def:A}) as special cases). This can be achieved by considering  the correlator 
\be\label{eq:Gcc}
G_{c\c}(\sigma)=\frac{\langle V(x_4) W(x_1) W(x_2) V(x_3)  \rangle}{\langle  W(x_1) W(x_2)   \rangle\langle V(x_4) V(x_3)  \rangle}
\ee
 in the kinematics (\ref{points}) in the dual CFT$_3$ description. The operators $W$ and $V$ are defined in (\ref{def:operators}) with $c,\c\in \mathbb{R}^N$. We have also introduced variables 
 \be\label{def:sigma}
 \eta=\rho\br\ , \qquad \sigma=\frac{1}{\r}\ .
 \ee
\subsection{Sum-Rules}
The correlator \eqref{eq:Gcc}, as explained in section \ref{sec:otoc}, can  be viewed as a thermal OTOC on Rindler space  
\be\label{definet_R}
G_{c\c}(\sigma)=F\(t_R-\frac{i\beta}{4}\)/F_d\ , \qquad e^{-2\pi t_R/\beta}=\sqrt{\eta}\sigma\ ,
\ee
where $t_R$ is the Rindler time. Analyticity of CFT correlators in Lorentzian signature, as discussed in \cite{Hartman:2016lgu} (see also \cite{Kundu:2019zsl,Kundu:2020gkz}), allows us to write a CFT$_3$ sum-rule for $A[c,\c]$
\be\label{CFTsumrule}
A[c,\c]=\frac{\tilde{\Delta}_f^4 \eta^{\frac{1}{2}}(1+Nc^2)(1+N\c^2)}{f_{3333}\(-\frac{1}{2}\log (\eta)\)}\lim_{\frac{1}{\Delta_f^4}\ll x\ll 1} \int^x_{0} d\sigma\ \mbox{Re}\(1-G_{c\c}(\sigma)\) \ge 0\ ,
\ee
for any $c,\c\in \mathbb{R}^N$ where $0<\eta<1$. Note that $f_{3333}$, as defined in (\ref{def:f}), is positive.\footnote{We have also absorbed positive numerical factors in the definition of $\tilde{\Delta}_f$.} Positivity of the integral follows from Rindler positivity which requires $\mbox{Re}\(1-G_{c\c}(\sigma)\) \ge 0$ \cite{Hartman:2016lgu}. This sum-rule  does not make any assumptions about the dual CFT$_3$ beyond the usual Euclidean axioms. Alternatively, the positivity also follows from the bound on the OTOC $|F(t_R-\frac{i\beta}{4})|\le F_d$, up to corrections that vanish in the limit $|\sigma| \ll 1$ \cite{Maldacena:2015waa}. Moreover, the above CFT$_3$ sum-rule, after using \eqref{definet_R}, can be  rewritten as a time integral of the OTOC 
\be\label{sumrule2}
A[c,\c]=\frac{1}{\beta}P(\eta) \lim_{t_*\gg t_0 \gg \beta} \int_{t_0}^{\infty} dt_R\ e^{-2\pi t_R/\beta}\ \mbox{Re}\(F_d-F\(t_R-\frac{i \beta}{4}\)\)\ge 0\ ,
\ee
where $P(\eta)$ is a theory independent positive function of $\eta$ that does not depend on $c$ and $\c$.\footnote{To be precise, 
\be
P(\eta)=\frac{512 \pi^5 \tilde{\Delta}_f^4 \eta^{3}}{9 f_{3333}\(-\frac{1}{2}\log (\eta)\)}\ , \qquad t_0=\frac{\beta}{2\pi}\log\(\frac{1}{\sqrt{\eta}x}\)\ .
\ee
Also note that the integral \eqref{sumrule2} does not depend on $t_0$ as long as it is much smaller than the effective scrambling time $t_*=\beta \log \(\Delta_f\) \gg t_0\gg \beta$. 
} 
A special case, of \eqref{sumrule2} is $c=\c=0$ which provides a relation equating $\Delta a$ with the integral of $F\(t_R-\frac{i \beta}{4}\)$. 

\subsection{$C$-Functions}
Another reason the sum-rules (\ref{CFTsumrule}) and \eqref{sumrule2} are of importance is that they provide a basis to construct an infinite set of CFT$_3$ functions that decrease monotonically along the RG flow  
\begin{align}\label{def:A}
A[\mu; c,\c]&=\frac{\tilde{\Delta}_f^4 \eta^{\frac{1}{2}}(1+Nc^2)(1+N\c^2)}{f_{3333}\(-\frac{1}{2}\log (\eta)\)}\lim_{\frac{1}{\Delta_f^4}\ll x\ll 1} \int^{x}_{\Delta_f^{-4\mu/f}x}  d\sigma\ \mbox{Re}\(1-G_{c\c}(\sigma)\)\nonumber\\
&=\frac{1}{\beta}P(\eta) \lim_{t_*\gg t_0 \gg \beta} \int_{t_0}^{t_0+ \frac{2\mu}{\pi f}t_*} dt_R\ e^{-\frac{2\pi t_R}{\beta}}\ \mbox{Re}\(F_d-F\(t_R-\frac{i \beta}{4}\)\)
\end{align}
with $0<\eta<1$, for all $c,\c\in \mathbb{R}^N$. These functions, for all $c,\c\in \mathbb{R}^N$, interpolate between $A[c,\c]$ in the UV ($\mu/f \rightarrow \infty$) and $0$ in the IR ($\mu/f\rightarrow 0$). Thus, using the basis (\ref{def:A}), we can construct a general $C$-function
\be\label{def:C}
C[\mu; c,\c]= C_{\rm IR}+\frac{C_{\rm UV}-C_{\rm IR}}{A[c,\c]}A[\mu; c,\c]\ .
\ee
By construction, $C[\mu; c,\c]$, for any $c,\c\in \mathbb{R}^N$, decreases monotonically from $C_{\rm UV}$ to $C_{\rm IR}$ under the RG flow. Besides,  $C[\mu; c,\c]$ defines a function which is constant and independent of energy scale at the UV and IR fixed points. A special case of (\ref{def:C}) with $C_{\rm UV}=a_{\rm UV}$ and $C_{\rm IR}=a_{\rm IR}$ is an infinite set of $a$-functions that monotonically decrease from $a_{\rm UV}$ to $a_{\rm IR}$. Any such $a$-function provides a good measure of the effective number of degrees of freedom along 4d RG flows.

Of course, in general (\ref{def:C}) is stronger than the special case we considered above. For example, it is possible that for certain values of $c$ and $\c$ the constraint \eqref{def:A1} for specific theories leads to a positivity condition for some other central charges associated with $\CFTUV$ and $\CFTIR$. For any such central charges, (\ref{def:C}) also provides a set of $C$-functions that interpolate between the UV and the IR values. 
  
Finally, let us comment on $C$-functions of 4d supersymmetric RG flows. The above discussion immediately implies that there are infinitely many distinct functions $a(\mu)$  for $\N=1$ supersymmetric flows that monotonically decrease along RG flows from $a_{\rm UV}$ to $a_{\rm IR}$. In particular, the CFT$_3$ quantity \eqref{def:C_susy} for any choice of $r_1$ and $r_2$ leads to  a distinct  $a(\mu)$.  

\section{Example: Free Massive Scalars}\label{sec:scalars}
The results of the preceding sections depend only on general principles and symmetries. We now provide a simple example that highlights most of the basic features of our general construction. 

Our UV theory contains a free complex scalar 
\be
\text{CFT}_{\rm UV}=-\frac{1}{2}\int d^4x  \p_\mu \Phi^\dagger \p^\mu \Phi \ , 
\ee
which enjoys an additional $U(1)$ global symmetry: $\Phi \rightarrow e^{i\theta}\Phi\ , \Phi^\dagger \rightarrow e^{-i\theta}\Phi^\dagger$. We now deform this CFT by adding mass terms 
\be\label{ex:uv}
S_{\rm UV}=-\frac{1}{2}\int d^4x  \(\p_\mu \Phi^\dagger \p^\mu \Phi +m_1^2 \Phi_1^2 +m_2^2 \Phi_2^2\)\ ,
\ee
where $\Phi=\Phi_1+i \Phi_2$ and $m_1^2,m_2^2>0$. The mass terms break both the conformal symmetry and the global $U(1)$ symmetry explicitly. We will discuss the flow of this UV theory. The $\CFTUV$ consists of two free massless scalars and hence 
\be\label{ex:auv}
a_{\rm UV}=2\times \frac{1}{360(4\pi)^2}\ .
\ee 
In the deep IR, the scalar field $\Phi$ decouples completely, and the $\CFTIR$ is trivial with no degrees of freedom, implying $a_{\rm IR}=0$. 

\subsection{Dilaton as a Compensator}
First as a warm up, we introduce a single real compensator $\Omega$ that allows us to view the explicit conformal symmetry breaking as a spontaneous symmetry breaking
\be\label{ex:mod}
S=-\frac{1}{2}\int d^4x  \(\p_\mu \Phi^\dagger \p^\mu \Phi+\p_\mu \Omega \p^\mu \Omega +\lambda_1 \Omega^2\Phi_1^2 +\lambda_2 \Omega^2 \Phi_2^2\)\ ,
\ee   
where $\lambda_1=m_1^2/f^2$ and $\lambda_2=m_2^2/f^2$ for some arbitrary mass scale $f$. The scale $f$ can be freely tuned, however, we do not want the compensator to modify the RG flow of (\ref{ex:uv}) and hence we choose $f\gg m_1,m_2$. In this limit, $\Omega$ and $\Phi$ are weakly interacting and we have perturbative control over the theory (\ref{ex:mod}).   

The theory (\ref{ex:mod}), at the classical level, is conformal. This can be seen by computing the classical stress tensor
\begin{align}
T_{\mu\nu}=&\partial_\mu \Phi_1 \partial_\nu \Phi_1+\partial_\mu \Phi_2 \partial_\nu \Phi_2+\partial_\mu \Omega \partial_\nu \Omega \nonumber\\
&-\frac{1}{2}\eta_{\mu\nu} \(\p_\mu \Phi^\dagger \p^\mu \Phi+\p_\mu \Omega \p^\mu \Omega +\lambda_1 \Omega^2\Phi_1^2 +\lambda_2 \Omega^2 \Phi_2^2\)
\end{align}
which is conserved but not traceless. This can be made traceless by adding an improvement term
\be
T^{\rm full}_{\mu \nu}=T_{\mu\nu}-\frac{1}{6} \left(\partial_\mu \partial_\nu-\eta_{\mu\nu}\Box \right)\left(\Phi \Phi^\dagger+\Omega^2 \right)\ .
\ee
The improved stress tensor $T^{\rm full}_{\mu \nu}$ is both conserved and traceless when we apply the equations of motion.

So, the conformal compensator modifies the theory (\ref{ex:uv}) into a classically conformal theory (\ref{ex:mod}). What does it imply for the quantum theory? Whenever conformal symmetry is broken explicitly by some mass parameters such as $m_1$ and $m_2$, that always introduces an operatorial anomaly to the trace of the stress tensor which spoils the anomaly matching argument of the previous sections. One can always introduce some conformal compensator $\Omega$ that removes the operatorial anomaly. This is reflected by the fact that $T^{\rm full}_{\mu \nu}$ is traceless. Moreover, the absence of the operatorial anomaly in (\ref{ex:mod}) guarantees that $a_{\rm UV}$ must match the total IR anomaly of $\CFTIR$ plus the dilaton. 

The theory (\ref{ex:mod}) has a moduli space along $\Omega$ for $\langle \Phi \rangle=0$. Clearly, the theory is conformal at $\langle \Omega \rangle=0$. However, the conformal symmetry is spontaneously broken at $\langle \Omega \rangle=f$ where we recover  (\ref{ex:uv}). Note that the theory (\ref{ex:mod}) does not have any global $U(1)$ symmetry even classically. From this perspective, the global $U(1)$ symmetry of $\CFTUV$ is emergent only at $\langle \Omega \rangle=0$. Hence, in this description we will not produce any NG boson for the broken $U(1)$ symmetry. 

The dilaton effective action can be obtained by studying fluctuations around the broken phase: $\Omega=f-\phi$. The action now becomes
\be
S=S_{\rm UV}-\frac{1}{2}\int d^4x  \(\p_\mu \phi \p^\mu \phi+J\(\frac{\phi}{f}\)\( m_1^2 \Phi_1^2 +m_2^2 \Phi_2^2\)\)\ , \quad J\(\frac{\phi}{f}\)=\frac{\phi^2}{f^2}-2 \frac{\phi}{f}\ .
\ee
At low energies, we can integrate out the massive fields $\Phi_1$ and $\Phi_2$. We proceed by computing the dilaton four-point amplitude at the leading order in $1/f$. This leads to precisely two copies of the 1-loop diagram in \cite{Komargodski:2011vj} for a single massive field -- one with $\Phi_1$ running in the loop and another one with $\Phi_2$ running in the loop. So, we get the following four-derivative effective action for the dilaton 
\be
S_{\rm conformal}[\phi]=-\frac{1}{2}\int d^4x \(\(\p \phi\)^2- \frac{1}{180(4\pi)^2 f^4}\(\phi^2 \Box^2 \phi^2\)+\O\(\phi^6; \p^6\)\)
\ee
which agrees with (\ref{eff:conformal}) for $\Delta a=a_{\rm UV}$ given by (\ref{ex:auv}).

\subsection{Spontaneous Breaking of $U(1)$ Symmetry}
Discussion of this section can be extended to also describe the explicit $U(1)$ symmetry breaking of (\ref{ex:uv}) as a spontaneous symmetry breaking. This can be done by introducing a complex compensator $\Omega$:
\be\label{ex:u1}
S=-\frac{1}{2}\int d^4x  \(\p_\mu \Phi^\dagger \p^\mu \Phi+\p_\mu \Omega \p^\mu \Omega^\dagger +\tilde{\lambda}_1 \Omega \Omega^\dagger \Phi \Phi^\dagger +\tilde{\lambda}_2 \(\Omega^2 {\Phi^\dagger}^2+{\Omega^\dagger}^2 \Phi^2\)\)\ ,
\ee
where couplings between $\Phi$ and $\Omega$ are arbitrarily weak 
\be
\tilde{\lambda}_1=\frac{m_1^2+m_2^2}{2f^2}\ , \qquad \tilde{\lambda}_2=\frac{m_1^2-m_2^2}{4f^2}\ .
\ee
Moreover, under the $U(1)$ symmetry $\Omega$ transforms as
\be
\Omega \rightarrow e^{i \theta} \Omega\ , \qquad \Omega^\dagger \rightarrow e^{-i\theta} \Omega^\dagger\ .
\ee
Similar to the previous case, the theory (\ref{ex:u1}) is classically conformal.  Furthermore, now one can also define a spin-1 current
\be
j_\mu=i\left(\Phi^\dagger \partial_\mu \Phi -\Phi \partial_\mu \Phi^\dagger \right)+i\left(\Omega^\dagger \partial_\mu \Omega -\Omega \partial_\mu \Omega^\dagger \right)
\ee
which is conserved, once we impose the equations of motion.

At the leading order in $\tilde{\lambda}_1$ and $\tilde{\lambda}_2$, the theory (\ref{ex:u1}) at energies $m_1,m_2\ll E\ll f$ can be viewed as an exactly conformal theory with a $U(1)$ global symmetry.  These symmetries are spontaneously broken when $\Omega$ gets a non-zero VEV, $\langle \Omega \rangle=f$, where we recover  (\ref{ex:uv}).
Fluctuations around the broken phase, $\Omega=f-\phi-i\xi$, create NG bosons associated with these broken symmetries. The additional massless mode $\xi$ arises from the spontaneous breaking of the global $U(1)$ symmetry.

The dilaton-axion effective action now can be obtained by integrating out the massive fields $\Phi_1$ and $\Phi_2$  from
\begin{align}
S=S_{\rm UV}-\frac{1}{2}\int d^4x & \(\p_\mu \phi \p^\mu \phi+\p_\mu \xi \p^\mu \xi+ J\(\frac{\phi}{f}\)\( m_1^2 \Phi_1^2 +m_2^2 \Phi_2^2\)\right. \nonumber\\
&\left. -8\tilde{\lambda}_2 \Phi_1\Phi_2 \(\phi\xi -f \xi\)+\frac{\xi^2}{f^2}\( m_2^2 \Phi_1^2 +m_1^2 \Phi_2^2\)\)\ .
\end{align}
As our general discussion led us to expect, the dilaton four-point scattering amplitude remains unchanged. In order to simplify the computations of other amplitudes, we take $m_2=m_1+\delta m$ with $\delta m \ll m_1$. In this limit, at the 1-loop level we find\footnote{Calculations of 1-loop scattering amplitudes of NG bosons in the context of free massive scalars can be performed in any spacetime dimensions following \cite{Elvang:2012yc}. }
\begin{align}
&\bA_4\(\phi\phi\phi\phi\) =  \frac{s^2+t^2+u^2}{45(4\pi)^2 f^4} \ ,\\
&\bA_4\( \xi\xi\xi\xi\) =\frac{s^2+t^2+u^2}{15(4\pi)^2 f^4} \(1+\frac{16}{3}\(\frac{\delta m}{m_1}\)^2+\O\(\frac{\delta m}{m_1}\)^4\)\ ,\\
&\bA_4\( \phi\phi\xi\xi\) =-\frac{s^2}{45(4\pi)^2 f^4}+\frac{ t^2+ u^2}{45(4\pi)^2 f^4}\(\frac{\delta m}{m_1}\)^2+\O\(\frac{\delta m}{m_1}\)^4
\end{align}
where $s=2p_1\cdot p_2$, $t=2p_1\cdot p_3$, and $u=2p_1\cdot p_4$. The resulting effective action has exactly the form (\ref{action}) with 
\be
B=6\Delta a\(1+\frac{16}{3}\(\frac{\delta m}{m_1}\)^2+\O\(\frac{\delta m}{m_1}\)^4\)\ , \qquad b=8\Delta a\(\frac{\delta m}{m_1}\)^2+\O\(\frac{\delta m}{m_1}\)^4\ .
\ee
These results are shown in figure \ref{intro:chaos}. Notice that $b=0$ only when $\Delta a=0$, which is consistent with the general results of section \ref{sec:u1}. Following our discussion of the preceding section, we can construct a set of functions that monotonically decreases from $a_{\rm UV}=\frac{1}{180(4\pi)^2}$ to $a_{\rm IR}=0$.

%
%%%%%%%%%%%%%%%%%%%%%%%%%%%%%
\section*{Acknowledgments}

It is my pleasure to thank Federico Bonetti, Diptarka Das, Jonathan Heckman, Jared Kaplan, Arnab Kundu, Tom Rudelius, and Hao Zhang for several helpful discussions. I am also grateful to Jonathan Heckman, Jared Kaplan, Arnab Kundu, and Hao Zhang for commenting on a draft. I was supported in part by the Simons Collaboration Grant on the Non-Perturbative Bootstrap.

\begin{appendix}
\section{Invariants for $U(1)$ Global Symmetry}\label{app:U1}
Four derivative invariants $W_i$ for $U(1)$ global symmetry are given by \cite{Bobev:2013vta}
\begin{align}\label{def_W}
&W_1=\hat{W}^2\ , \qquad W_2=\hat{R}^2\ , \qquad W_3=\hat{A}_\mu \hat{\nabla}^\mu \hat{R}\ , \qquad W_4=\( \hat{\nabla}^\mu \hat{A}_\mu\)^2\ ,\\
&W_5=\hat{g}^{\mu\nu}\hat{A}_\mu \hat{\Box}\hat{A}_\nu \ , \qquad W_6=\hat{R}^{\mu\nu}\hat{A}_\mu\hat{A}_\nu\ , \qquad W_7=\hat{R}\hat{g}^{\mu\nu}\hat{A}_\mu\hat{A}_\nu\ , \qquad W_8=\(\hat{g}^{\mu\nu}\hat{A}_\mu\hat{A}_\nu\)^2\ , \nonumber\\
&W_9=\hat{g}^{\mu\nu}\hat{A}_\mu\hat{A}_\nu \hat{\nabla}^\lambda \hat{A}_\lambda\ , \nonumber
\end{align}
where, $\hat{R}$, $\hat{R}^{\mu\nu}$, and $\hat{W}_{\mu\nu\alpha\beta}$ are computed using the Weyl-invariant metric (\ref{metric0}). Terms that vanish once we impose the on-shell condition for the dilaton and the axion can be safely ignored at low energies since these terms can only affect low energy observables at subleading orders. Hence, the Weyl invariants $W_i$ in the flat space limit with no background gauge field can be further simplified by using the free equations of motion (\ref{eq:eom})
\begin{align}\label{def_W1}
&W_1=0\ , \qquad W_2=36 \gamma_0^4 e^{4\tau}\(\p \beta\)^4\ , \qquad W_3=0\ , \qquad W_4=0\ ,\\
&W_5= W_6=-2e^{4\tau}\(\gamma_0^2\(\p \beta\)^4+\(\p \tau \cdot \p \beta\)^2\)\ , \qquad W_7=-6\gamma_0^2 e^{4\tau}\(\p \beta\)^4\ ,   \nonumber\\
&W_8=e^{4\tau}\(\p \beta\)^4\ ,\qquad W_9=0\ . \nonumber
\end{align}

\section{The Effective Action for Non-Abelian Global Symmetries}\label{app:NA}

\subsection{Invariants for Non-Abelian Global Symmetries}
Independent four derivative invariants $\tilde{W}_i$ for a general compact, simple Lie group $G$ are given by\footnote{Note that there can be Wess-Zumino type terms  that are not exactly invariant but shifts by a total derivative under  the gauge and Weyl transformations. For a detailed discussion see \cite{DHoker:1994rdl}. However, these terms do not contribute at the 4-field 4-derivative level and hence we will ignore them.}
\begin{align}\label{def_W2}
&\tilde{W}_1=\Tr\(\(\hat{\nabla}^\mu \hat{A}_\mu\)\( \hat{\nabla}^\nu \hat{A}_\nu\)\) \ ,\qquad \tilde{W}_2=\hat{g}^{\mu\nu}\Tr\(\hat{A}_\mu \hat{\Box}\hat{A}_\nu\)\ , \\
&\tilde{W}_3=\hat{R}^{\mu\nu}\Tr\(\hat{A}_\mu\hat{A}_\nu\)\ , \qquad \tilde{W}_4=\hat{R}\hat{g}^{\mu\nu}\Tr\(\hat{A}_\mu\hat{A}_\nu\)\ , \qquad \tilde{W}_5=\hat{g}^{\mu\nu}\Tr\(\hat{A}_\mu\hat{A}_\rho \hat{\nabla}^\rho \hat{A}_\nu\)\ ,\nonumber\\
&\tilde{W}_6=\hat{g}^{\mu\nu}\hat{g}^{\rho\sigma}\Tr\(\hat{A}_\mu\hat{A}_\nu\)\Tr\(\hat{A}_\rho\hat{A}_\sigma\)\ , \qquad \tilde{W}_7=\hat{g}^{\mu\nu}\hat{g}^{\rho\sigma}\Tr\(\hat{A}_\mu\hat{A}_\rho\)\Tr\(\hat{A}_\nu\hat{A}_\sigma\)\ ,  \nonumber\\
&\tilde{W}_8=\hat{g}^{\mu\nu}\hat{g}^{\rho\sigma}\Tr\(\hat{A}_\mu \hat{A}_\nu\hat{A}_\rho \hat{A}_\sigma\)\ ,\qquad \tilde{W}_{9}=\hat{W}^2\ , \qquad \tilde{W}_{10}=\hat{R}^2 \ .\nonumber
\end{align}
We now write down the low-energy effective action by taking the flat space limit with no background gauge field of \eqref{inv}
\begin{align}
S_{\text{eff}}[\tau,\beta]&= \int d^4x \(-\frac{f^2}{2}e^{-2\tau}\left(\(\p \tau\)^2+\sum_{i=1}^N \(D \beta_i\)^2\)+2\Delta a \(\p \tau\)^2 \(2\Box \tau-\(\p \tau\)^2\)\)\nonumber \\
&+\int d^4x\ e^{-4\tau}\left(\sum_{I=1}^{10} \gamma_I \tilde{W}_I\)_{g_{\mu\nu}=\eta_{\mu\nu}, A_\mu=0}+\cdots\ ,
\end{align}
where dots represent higher derivative terms. Note that we removed $\gamma_0$, which is theory dependent, by rescaling the generators
\be\label{norm}
\Tr\(T_i T_j\)=\frac{1}{2\gamma_0^2} \delta_{ij}\ .
\ee
The Maurer-Cartan form $D_\mu \beta$ is given by
\begin{align}
D_\mu \beta_i=\p_\mu \beta_i-\frac{1}{2}f_{ijk} \beta_j \p_\mu \beta_k +\frac{1}{6}f_{njk}f_{nil}\beta_l \beta_j \p_\mu \beta_k+\cdots \ .
\end{align}
Equations of motion at the leading order are 
\be
\Box \tau=\(\p \tau\)^2-\sum_{i=1}^N \(\p \beta_i\)^2+\cdots\ , \qquad \Box \beta_i=2\(\p \tau \cdot \p \beta_i\)+\cdots\ .
\ee
The invariants $\tilde{W}_I$ in the flat space limit with no background gauge field can be further simplified by using the above equations of motion. At the four-field level we obtain
\begin{align}
&\tilde{W}_1=0\ , \qquad \tilde{W}_2=\tilde{W}_3=-e^{4\tau}\(\(\sum_i (\p\beta_i)^2\)^2+\delta_{ij}\(\p \tau \cdot \p \beta_i\)\(\p \tau \cdot \p \beta_j\)\)\  \ ,\\
&\tilde{W}_4=-3e^{4\tau}\(\sum_i (\p\beta_i)^2\)^2\ ,  \qquad \tilde{W}_5=e^{4\tau}f_{ijk}f_{ij'k'} \(\p \beta_j\cdot \p \beta_{j'}\)\(\p \beta_k\cdot \p \beta_{k'}\)\ , \nonumber\\
&\tilde{W}_6=\frac{1}{4}e^{4\tau}\(\sum_i (\p\beta_i)^2\)^2\ , \qquad \tilde{W}_7=\frac{1}{4}e^{4\tau}\(\p \beta_j\cdot \p \beta_{k}\)\(\p \beta_j\cdot \p \beta_{k}\)\ ,\nonumber\\
&\tilde{W}_8=T_{ijkl}e^{4\tau}\(\p \beta_i\cdot \p \beta_j\)\(\p \beta_k\cdot \p \beta_l\)\ , \qquad T_{ijkl}=\Tr\(\{T_i, T_j\}\{T_k,T_l\}\)\ ,\nonumber\\
& \tilde{W}_{9}=0\ , \qquad \tilde{W}_{10}=36 e^{4\tau} \(\sum_i (\p\beta_i)^2\)^2\ .\nonumber
\end{align}
Note that there is another possible invariant $\hat{g}^{\mu\nu}\hat{g}^{\rho\sigma}\Tr\(\hat{A}_\mu \hat{A}_\rho \hat{A}_\nu \hat{A}_\sigma\)$, however, this term at the four-field four-derivative level can be expressed as a linear combination of $ \tilde{W}_5$ and $ \tilde{W}_8$. Moreover, one can also construct a parity odd invariant $\epsilon^{\mu\nu\rho\sigma}\Tr\(\hat{A}_\mu \hat{A}_\nu\hat{A}_\rho \hat{A}_\sigma\)$ which is a total derivative. 

\subsection{Effective Action}
We can now write the effective action at the four-derivative four-field level
\begin{align}\label{semifinal}
S_{\text{eff}}[\tau,\beta_i]=& -\frac{f^2}{2}\int d^4x e^{-2\tau}\left(\(\p \tau\)^2+ \sum_{i=1}^N \(\p \beta_i\)^2-\frac{1}{12}f_{ijk}f_{ij'k'} \beta_j \beta_{j'}\(\p \beta_k \cdot \p \beta_{k'}\)+\cdots\)\nonumber\\
&+2\Delta a  \int d^4x \(\(\p \tau\)^4 - 2\(\p \tau\)^2 \sum_{i=1}^N \(\p \beta_i\)^2\) \\
&+\int d^4x \( b \(\p \tau \cdot \p \beta_i\)\(\p \tau \cdot \p \beta_i\)+  B_{ijkl}\(\p \beta_i\cdot \p \beta_j\)\(\p \beta_k\cdot \p \beta_l\)\) +\cdots\ ,\nonumber
\end{align}
where, indices $i,j,\cdots\in\{1,2,\cdots,N=\mbox{dim}\ G\}$. Coefficients $B_{ijkl}$ satisfy
\be
B_{ijkl}=B_1 \delta_{ij}\delta_{kl}+B_2 \(\delta_{ik}\delta_{jl}+\delta_{il}\delta_{jk}\)+B_3\( f_{i'ik}f_{i'jl}+f_{i'il}f_{i'jk}\)+B_4T_{ijkl}\ ,
\ee
where, $b,B_1,B_2,B_3,B_4$ are arbitrary coefficients and  $T_{ijkl}=\Tr\(\{T_i, T_j\}\{T_k,T_l\}\)$, $f_{ijk}=-2i \Tr\([T^i,T^j]T^k\)$. Note that for $G=U(1)$, this action agrees with (\ref{simple}). 

Let us now write the effective action (\ref{semifinal}) in a more traditional form by performing the following field redefinition: 
\begin{align}
e^{-\tau} \sin \beta_i= \frac{\xi_i }{f}\ , \qquad e^{-\tau}=\sqrt{\(1-\frac{\phi}{f}\)^2+\frac{\xi^2}{f^2}}\ ,
\end{align}
where $\xi^2=\xi_i \xi_i$. In terms of the physical fields $\phi$ and $\xi_i$, the effective action at the four-derivative and four-field level can be written as
\begin{align}\label{app:final}
S_{\text{eff}}[\phi,\xi_i]=& -\frac{1}{2}\int d^4x \left((\p \phi)^2+  \p \xi_i\cdot \p \xi_i-\frac{1}{12f^2 }f_{ijk}f_{ij'k'} \xi_j \xi_{j'}\(\p \xi_k \cdot \p \xi_{k'}\)+ \frac{1}{4 f^2}\sum_{i\neq j} \xi_i^2 \Box \xi_j^2+\cdots\)\nonumber\\
&+\frac{2\Delta a }{f^4} \int d^4x \(\(\p \phi\)^4 - 2\(\p \phi\)^2  \(\p \xi_i\cdot \p \xi_i\)\) \\
&+\frac{1}{f^4}\int d^4x \( b \(\p \phi \cdot \p \xi_i\)\(\p \phi \cdot \p \xi_i\)+  B_{ijkl}\(\p \xi_i\cdot \p \xi_j\)\(\p \xi_k\cdot \p \xi_l\)\) +\cdots\ .\nonumber
\end{align}

\end{appendix}

%%%%%%%%%%%%%%%%%%%%%%%%%%%%%%%%%%%%%%%%%%%

\end{spacing}

\bibliographystyle{utphys} 
\bibliography{RGBib}

\providecommand{\href}[2]{#2}\begingroup\raggedright\begin{thebibliography}{10}

\bibitem{Maldacena:2015waa}
J.~Maldacena, S.~H. Shenker, and D.~Stanford, ``{A bound on chaos},''
\href{http://arxiv.org/abs/1503.01409}{{\ttfamily arXiv:1503.01409 [hep-th]}}.
%%CITATION = ARXIV:1503.01409;%%.

\bibitem{Cordova:2015fha}
C.~Cordova, T.~T. Dumitrescu, and K.~Intriligator, ``{Anomalies,
  renormalization group flows, and the a-theorem in six-dimensional (1, 0)
  theories},'' \href{http://dx.doi.org/10.1007/JHEP10(2016)080}{{\em JHEP}
  {\bfseries 10} (2016) 080}, \href{http://arxiv.org/abs/1506.03807}{{\ttfamily
  arXiv:1506.03807 [hep-th]}}.

\bibitem{Zamolodchikov:1986gt}
A.~Zamolodchikov, ``{Irreversibility of the Flux of the Renormalization Group
  in a 2D Field Theory},'' {\em JETP Lett.} {\bfseries 43} (1986) 730--732.

\bibitem{Komargodski:2011vj}
Z.~Komargodski and A.~Schwimmer, ``{On Renormalization Group Flows in Four
  Dimensions},'' \href{http://dx.doi.org/10.1007/JHEP12(2011)099}{{\em JHEP}
  {\bfseries 1112} (2011) 099},
\href{http://arxiv.org/abs/1107.3987}{{\ttfamily arXiv:1107.3987 [hep-th]}}.
%%CITATION = ARXIV:1107.3987;%%.

\bibitem{Komargodski:2011xv}
Z.~Komargodski, ``{The Constraints of Conformal Symmetry on RG Flows},''
  \href{http://dx.doi.org/10.1007/JHEP07(2012)069}{{\em JHEP} {\bfseries 07}
  (2012) 069}, \href{http://arxiv.org/abs/1112.4538}{{\ttfamily arXiv:1112.4538
  [hep-th]}}.

\bibitem{Cardy:1988cwa}
J.~L. Cardy, ``{Is There a c Theorem in Four-Dimensions?},''
  \href{http://dx.doi.org/10.1016/0370-2693(88)90054-8}{{\em Phys. Lett. B}
  {\bfseries 215} (1988) 749--752}.

\bibitem{Elvang:2012st}
H.~Elvang, D.~Z. Freedman, L.-Y. Hung, M.~Kiermaier, R.~C. Myers, and
  S.~Theisen, ``{On renormalization group flows and the a-theorem in 6d},''
  \href{http://dx.doi.org/10.1007/JHEP10(2012)011}{{\em JHEP} {\bfseries 10}
  (2012) 011}, \href{http://arxiv.org/abs/1205.3994}{{\ttfamily arXiv:1205.3994
  [hep-th]}}.

\bibitem{Kundu:2019zsl}
S.~Kundu, ``{Renormalization Group Flows, the $a$-Theorem and Conformal
  Bootstrap},'' \href{http://dx.doi.org/10.1007/JHEP05(2020)014}{{\em JHEP}
  {\bfseries 05} (2020) 014}, \href{http://arxiv.org/abs/1912.09479}{{\ttfamily
  arXiv:1912.09479 [hep-th]}}.

\bibitem{Cordova:2015vwa}
C.~Cordova, T.~T. Dumitrescu, and X.~Yin, ``{Higher derivative terms, toroidal
  compactification, and Weyl anomalies in six-dimensional (2, 0) theories},''
  \href{http://dx.doi.org/10.1007/JHEP10(2019)128}{{\em JHEP} {\bfseries 10}
  (2019) 128}, \href{http://arxiv.org/abs/1505.03850}{{\ttfamily
  arXiv:1505.03850 [hep-th]}}.

\bibitem{Heckman:2015axa}
J.~J. Heckman and T.~Rudelius, ``{Evidence for C-theorems in 6D SCFTs},''
  \href{http://dx.doi.org/10.1007/JHEP09(2015)218}{{\em JHEP} {\bfseries 09}
  (2015) 218}, \href{http://arxiv.org/abs/1506.06753}{{\ttfamily
  arXiv:1506.06753 [hep-th]}}.

\bibitem{Heckman:2016ssk}
J.~J. Heckman, T.~Rudelius, and A.~Tomasiello, ``{6D RG Flows and Nilpotent
  Hierarchies},'' \href{http://dx.doi.org/10.1007/JHEP07(2016)082}{{\em JHEP}
  {\bfseries 07} (2016) 082}, \href{http://arxiv.org/abs/1601.04078}{{\ttfamily
  arXiv:1601.04078 [hep-th]}}.

\bibitem{Heckman:2018jxk}
J.~J. Heckman and T.~Rudelius, ``{Top Down Approach to 6D SCFTs},''
  \href{http://dx.doi.org/10.1088/1751-8121/aafc81}{{\em J. Phys. A} {\bfseries
  52} no.~9, (2019) 093001}, \href{http://arxiv.org/abs/1805.06467}{{\ttfamily
  arXiv:1805.06467 [hep-th]}}.

\bibitem{Afkhami-Jeddi:2016ntf}
N.~Afkhami-Jeddi, T.~Hartman, S.~Kundu, and A.~Tajdini, ``{Einstein gravity
  3-point functions from conformal field theory},''
  \href{http://dx.doi.org/10.1007/JHEP12(2017)049}{{\em JHEP} {\bfseries 12}
  (2017) 049},
\href{http://arxiv.org/abs/1610.09378}{{\ttfamily arXiv:1610.09378 [hep-th]}}.
%%CITATION = ARXIV:1610.09378;%%.

\bibitem{Kundu:2020gkz}
S.~Kundu, ``{A Generalized Nachtmann Theorem in CFT},''
  \href{http://dx.doi.org/10.1007/JHEP11(2020)138}{{\em JHEP} {\bfseries 11}
  (2020) 138}, \href{http://arxiv.org/abs/2002.12390}{{\ttfamily
  arXiv:2002.12390 [hep-th]}}.

\bibitem{Bobev:2013vta}
N.~Bobev, H.~Elvang, and T.~M. Olson, ``{Dilaton effective action with N = 1
  supersymmetry},'' \href{http://dx.doi.org/10.1007/JHEP04(2014)157}{{\em JHEP}
  {\bfseries 04} (2014) 157}, \href{http://arxiv.org/abs/1312.2925}{{\ttfamily
  arXiv:1312.2925 [hep-th]}}.

\bibitem{Heemskerk:2009pn}
I.~Heemskerk, J.~Penedones, J.~Polchinski, and J.~Sully, ``{Holography from
  Conformal Field Theory},''
  \href{http://dx.doi.org/10.1088/1126-6708/2009/10/079}{{\em JHEP} {\bfseries
  0910} (2009) 079},
\href{http://arxiv.org/abs/0907.0151}{{\ttfamily arXiv:0907.0151 [hep-th]}}.
%%CITATION = ARXIV:0907.0151;%%.

\bibitem{Heemskerk:2010ty}
I.~Heemskerk and J.~Sully, ``{More Holography from Conformal Field Theory},''
  \href{http://dx.doi.org/10.1007/JHEP09(2010)099}{{\em JHEP} {\bfseries 1009}
  (2010) 099},
\href{http://arxiv.org/abs/1006.0976}{{\ttfamily arXiv:1006.0976 [hep-th]}}.
%%CITATION = ARXIV:1006.0976;%%.

\bibitem{Fitzpatrick:2010zm}
A.~Fitzpatrick, E.~Katz, D.~Poland, and D.~Simmons-Duffin, ``{Effective
  Conformal Theory and the Flat-Space Limit of AdS},''
  \href{http://dx.doi.org/10.1007/JHEP07(2011)023}{{\em JHEP} {\bfseries 1107}
  (2011) 023},
\href{http://arxiv.org/abs/1007.2412}{{\ttfamily arXiv:1007.2412 [hep-th]}}.
%%CITATION = ARXIV:1007.2412;%%.

\bibitem{Penedones:2010ue}
J.~Penedones, ``{Writing CFT correlation functions as AdS scattering
  amplitudes},'' \href{http://dx.doi.org/10.1007/JHEP03(2011)025}{{\em JHEP}
  {\bfseries 1103} (2011) 025},
\href{http://arxiv.org/abs/1011.1485}{{\ttfamily arXiv:1011.1485 [hep-th]}}.
%%CITATION = ARXIV:1011.1485;%%.

\bibitem{ElShowk:2011ag}
S.~El-Showk and K.~Papadodimas, ``{Emergent Spacetime and Holographic CFTs},''
  \href{http://dx.doi.org/10.1007/JHEP10(2012)106}{{\em JHEP} {\bfseries 1210}
  (2012) 106},
\href{http://arxiv.org/abs/1101.4163}{{\ttfamily arXiv:1101.4163 [hep-th]}}.
%%CITATION = ARXIV:1101.4163;%%.

\bibitem{Fitzpatrick:2011ia}
A.~L. Fitzpatrick, J.~Kaplan, J.~Penedones, S.~Raju, and B.~C. van Rees, ``{A
  Natural Language for AdS/CFT Correlators},''
  \href{http://dx.doi.org/10.1007/JHEP11(2011)095}{{\em JHEP} {\bfseries 1111}
  (2011) 095},
\href{http://arxiv.org/abs/1107.1499}{{\ttfamily arXiv:1107.1499 [hep-th]}}.
%%CITATION = ARXIV:1107.1499;%%.

\bibitem{Fitzpatrick:2011hu}
A.~L. Fitzpatrick and J.~Kaplan, ``{Analyticity and the Holographic
  S-Matrix},'' \href{http://dx.doi.org/10.1007/JHEP10(2012)127}{{\em JHEP}
  {\bfseries 1210} (2012) 127},
\href{http://arxiv.org/abs/1111.6972}{{\ttfamily arXiv:1111.6972 [hep-th]}}.
%%CITATION = ARXIV:1111.6972;%%.

\bibitem{Fitzpatrick:2011dm}
A.~L. Fitzpatrick and J.~Kaplan, ``{Unitarity and the Holographic S-Matrix},''
  \href{http://dx.doi.org/10.1007/JHEP10(2012)032}{{\em JHEP} {\bfseries 1210}
  (2012) 032},
\href{http://arxiv.org/abs/1112.4845}{{\ttfamily arXiv:1112.4845 [hep-th]}}.
%%CITATION = ARXIV:1112.4845;%%.

\bibitem{Fitzpatrick:2012cg}
A.~Fitzpatrick and J.~Kaplan, ``{AdS Field Theory from Conformal Field
  Theory},'' \href{http://dx.doi.org/10.1007/JHEP02(2013)054}{{\em JHEP}
  {\bfseries 02} (2013) 054}, \href{http://arxiv.org/abs/1208.0337}{{\ttfamily
  arXiv:1208.0337 [hep-th]}}.

\bibitem{Goncalves:2014rfa}
V.~Goncalves, J.~Penedones, and E.~Trevisani, ``{Factorization of Mellin
  amplitudes},'' \href{http://dx.doi.org/10.1007/JHEP10(2015)040}{{\em JHEP}
  {\bfseries 10} (2015) 040}, \href{http://arxiv.org/abs/1410.4185}{{\ttfamily
  arXiv:1410.4185 [hep-th]}}.

\bibitem{Alday:2014tsa}
L.~F. Alday, A.~Bissi, and T.~Lukowski, ``{Lessons from crossing symmetry at
  large N},''
\href{http://arxiv.org/abs/1410.4717}{{\ttfamily arXiv:1410.4717 [hep-th]}}.
%%CITATION = ARXIV:1410.4717;%%.

\bibitem{Hijano:2015zsa}
E.~Hijano, P.~Kraus, E.~Perlmutter, and R.~Snively, ``{Witten Diagrams
  Revisited: The AdS Geometry of Conformal Blocks},''
\href{http://arxiv.org/abs/1508.00501}{{\ttfamily arXiv:1508.00501 [hep-th]}}.
%%CITATION = ARXIV:1508.00501;%%.

\bibitem{Aharony:2016dwx}
O.~Aharony, L.~F. Alday, A.~Bissi, and E.~Perlmutter, ``{Loops in AdS from
  Conformal Field Theory},''
  \href{http://dx.doi.org/10.1007/JHEP07(2017)036}{{\em JHEP} {\bfseries 07}
  (2017) 036}, \href{http://arxiv.org/abs/1612.03891}{{\ttfamily
  arXiv:1612.03891 [hep-th]}}.

\bibitem{Schwimmer:2010za}
A.~Schwimmer and S.~Theisen, ``{Spontaneous Breaking of Conformal Invariance
  and Trace Anomaly Matching},''
  \href{http://dx.doi.org/10.1016/j.nuclphysb.2011.02.003}{{\em Nucl. Phys. B}
  {\bfseries 847} (2011) 590--611},
  \href{http://arxiv.org/abs/1011.0696}{{\ttfamily arXiv:1011.0696 [hep-th]}}.

\bibitem{Remmen:2019cyz}
G.~N. Remmen and N.~L. Rodd, ``{Consistency of the Standard Model Effective
  Field Theory},'' \href{http://dx.doi.org/10.1007/JHEP12(2019)032}{{\em JHEP}
  {\bfseries 12} (2019) 032}, \href{http://arxiv.org/abs/1908.09845}{{\ttfamily
  arXiv:1908.09845 [hep-ph]}}.

\bibitem{Bellazzini:2020cot}
B.~Bellazzini, J.~Elias~Mir\'o, R.~Rattazzi, M.~Riembau, and F.~Riva,
  ``{Positive Moments for Scattering Amplitudes},''
  \href{http://arxiv.org/abs/2011.00037}{{\ttfamily arXiv:2011.00037
  [hep-th]}}.

\bibitem{Tolley:2020gtv}
A.~J. Tolley, Z.-Y. Wang, and S.-Y. Zhou, ``{New positivity bounds from full
  crossing symmetry},'' \href{http://arxiv.org/abs/2011.02400}{{\ttfamily
  arXiv:2011.02400 [hep-th]}}.

\bibitem{Trott:2020ebl}
T.~Trott, ``{Causality, Unitarity and Symmetry in Effective Field Theory},''
  \href{http://arxiv.org/abs/2011.10058}{{\ttfamily arXiv:2011.10058
  [hep-ph]}}.

\bibitem{Paulos:2016fap}
M.~F. Paulos, J.~Penedones, J.~Toledo, B.~C. van Rees, and P.~Vieira, ``{The
  S-matrix Bootstrap I: QFT in AdS},''
\href{http://arxiv.org/abs/1607.06109}{{\ttfamily arXiv:1607.06109 [hep-th]}}.
%%CITATION = ARXIV:1607.06109;%%.

\bibitem{Paulos:2016but}
M.~F. Paulos, J.~Penedones, J.~Toledo, B.~C. van Rees, and P.~Vieira, ``{The
  S-matrix bootstrap II: two dimensional amplitudes},''
  \href{http://dx.doi.org/10.1007/JHEP11(2017)143}{{\em JHEP} {\bfseries 11}
  (2017) 143}, \href{http://arxiv.org/abs/1607.06110}{{\ttfamily
  arXiv:1607.06110 [hep-th]}}.

\bibitem{Paulos:2017fhb}
M.~F. Paulos, J.~Penedones, J.~Toledo, B.~C. van Rees, and P.~Vieira, ``{The
  S-matrix bootstrap. Part III: higher dimensional amplitudes},''
  \href{http://dx.doi.org/10.1007/JHEP12(2019)040}{{\em JHEP} {\bfseries 12}
  (2019) 040}, \href{http://arxiv.org/abs/1708.06765}{{\ttfamily
  arXiv:1708.06765 [hep-th]}}.

\bibitem{Homrich:2019cbt}
A.~Homrich, J.~Penedones, J.~Toledo, B.~C. van Rees, and P.~Vieira, ``{The
  S-matrix Bootstrap IV: Multiple Amplitudes},''
  \href{http://dx.doi.org/10.1007/JHEP11(2019)076}{{\em JHEP} {\bfseries 11}
  (2019) 076}, \href{http://arxiv.org/abs/1905.06905}{{\ttfamily
  arXiv:1905.06905 [hep-th]}}.

\bibitem{1969JETP...28.1200L}
A.~I. {Larkin} and Y.~N. {Ovchinnikov}, ``{Quasiclassical Method in the Theory
  of Superconductivity},'' {\em Soviet Journal of Experimental and Theoretical
  Physics} {\bfseries 28} (June, 1969) 1200.

\bibitem{kitaev2014chaos}
A.~Kitaev, ``Hidden correlations in the hawking radiation and thermal noise,''
  in {\em https://youtu.be/OQ9qN8j7EZI}.
\newblock 2014.

\bibitem{Hartman:2016lgu}
T.~Hartman, S.~Kundu, and A.~Tajdini, ``{Averaged Null Energy Condition from
  Causality},'' \href{http://dx.doi.org/10.1007/JHEP07(2017)066}{{\em JHEP}
  {\bfseries 07} (2017) 066},
\href{http://arxiv.org/abs/1610.05308}{{\ttfamily arXiv:1610.05308 [hep-th]}}.
%%CITATION = ARXIV:1610.05308;%%.

\bibitem{Kundu:2020nir}
A.~Kundu, ``{Fast Scrambling under an RG-flow},''
  \href{http://arxiv.org/abs/2006.11037}{{\ttfamily arXiv:2006.11037
  [hep-th]}}.

\bibitem{Adams:2006sv}
A.~Adams, N.~Arkani-Hamed, S.~Dubovsky, A.~Nicolis, and R.~Rattazzi,
  ``{Causality, analyticity and an IR obstruction to UV completion},''
  \href{http://dx.doi.org/10.1088/1126-6708/2006/10/014}{{\em JHEP} {\bfseries
  0610} (2006) 014},
\href{http://arxiv.org/abs/hep-th/0602178}{{\ttfamily arXiv:hep-th/0602178
  [hep-th]}}.
%%CITATION = HEP-TH/0602178;%%.

\bibitem{Myers:2010xs}
R.~C. Myers and A.~Sinha, ``{Seeing a c-theorem with holography},''
  \href{http://dx.doi.org/10.1103/PhysRevD.82.046006}{{\em Phys. Rev. D}
  {\bfseries 82} (2010) 046006},
  \href{http://arxiv.org/abs/1006.1263}{{\ttfamily arXiv:1006.1263 [hep-th]}}.

\bibitem{Myers:2010tj}
R.~C. Myers and A.~Sinha, ``{Holographic c-theorems in arbitrary dimensions},''
  \href{http://dx.doi.org/10.1007/JHEP01(2011)125}{{\em JHEP} {\bfseries 01}
  (2011) 125},
\href{http://arxiv.org/abs/1011.5819}{{\ttfamily arXiv:1011.5819 [hep-th]}}.
%%CITATION = ARXIV:1011.5819;%%.

\bibitem{Hofman:2008ar}
D.~M. Hofman and J.~Maldacena, ``{Conformal collider physics: Energy and charge
  correlations},'' \href{http://dx.doi.org/10.1088/1126-6708/2008/05/012}{{\em
  JHEP} {\bfseries 0805} (2008) 012},
\href{http://arxiv.org/abs/0803.1467}{{\ttfamily arXiv:0803.1467 [hep-th]}}.
%%CITATION = ARXIV:0803.1467;%%.

\bibitem{Komargodski:2016gci}
Z.~Komargodski, M.~Kulaxizi, A.~Parnachev, and A.~Zhiboedov, ``{Conformal Field
  Theories and Deep Inelastic Scattering},''
  \href{http://dx.doi.org/10.1103/PhysRevD.95.065011}{{\em Phys. Rev. D}
  {\bfseries 95} no.~6, (2017) 065011},
  \href{http://arxiv.org/abs/1601.05453}{{\ttfamily arXiv:1601.05453
  [hep-th]}}.

\bibitem{Hartman:2016dxc}
T.~Hartman, S.~Jain, and S.~Kundu, ``{A New Spin on Causality Constraints},''
  \href{http://dx.doi.org/10.1007/JHEP10(2016)141}{{\em JHEP} {\bfseries 10}
  (2016) 141},
\href{http://arxiv.org/abs/1601.07904}{{\ttfamily arXiv:1601.07904 [hep-th]}}.
%%CITATION = ARXIV:1601.07904;%%.

\bibitem{Hofman:2016awc}
D.~M. Hofman, D.~Li, D.~Meltzer, D.~Poland, and F.~Rejon-Barrera, ``{A Proof of
  the Conformal Collider Bounds},''
  \href{http://dx.doi.org/10.1007/JHEP06(2016)111}{{\em JHEP} {\bfseries 06}
  (2016) 111},
\href{http://arxiv.org/abs/1603.03771}{{\ttfamily arXiv:1603.03771 [hep-th]}}.
%%CITATION = ARXIV:1603.03771;%%.

\bibitem{Radicevic:2016kpf}
D.~Radicevic, ``{Quantum Mechanics in the Infrared},''
  \href{http://arxiv.org/abs/1608.07275}{{\ttfamily arXiv:1608.07275
  [hep-th]}}.

\bibitem{Srednicki:1995pt}
M.~Srednicki, ``{Thermal fluctuations in quantized chaotic systems},''
  \href{http://dx.doi.org/10.1088/0305-4470/29/4/003}{{\em J. Phys. A}
  {\bfseries 29} (1996) L75--L79},
  \href{http://arxiv.org/abs/chao-dyn/9511001}{{\ttfamily
  arXiv:chao-dyn/9511001}}.

\bibitem{PhysRevA.43.2046}
J.~M. Deutsch, ``Quantum statistical mechanics in a closed system,''
  \href{http://dx.doi.org/10.1103/PhysRevA.43.2046}{{\em Phys. Rev. A}
  {\bfseries 43} (Feb, 1991) 2046--2049}.
  \url{https://link.aps.org/doi/10.1103/PhysRevA.43.2046}.

\bibitem{2008Natur.452..854R}
M.~{Rigol}, V.~{Dunjko}, and M.~{Olshanii}, ``{Thermalization and its mechanism
  for generic isolated quantum systems},''
  \href{http://dx.doi.org/10.1038/nature06838}{{\em Nature} {\bfseries 452}
  no.~7189, (Apr, 2008) 854--858},
  \href{http://arxiv.org/abs/0708.1324}{{\ttfamily arXiv:0708.1324
  [cond-mat.stat-mech]}}.

\bibitem{Coleman:1969sm}
S.~R. Coleman, J.~Wess, and B.~Zumino, ``{Structure of phenomenological
  Lagrangians. 1.},'' \href{http://dx.doi.org/10.1103/PhysRev.177.2239}{{\em
  Phys. Rev.} {\bfseries 177} (1969) 2239--2247}.

\bibitem{Callan:1969sn}
J.~Callan, Curtis~G., S.~R. Coleman, J.~Wess, and B.~Zumino, ``{Structure of
  phenomenological Lagrangians. 2.},''
  \href{http://dx.doi.org/10.1103/PhysRev.177.2247}{{\em Phys. Rev.} {\bfseries
  177} (1969) 2247--2250}.

\bibitem{Weinberg:1996kr}
S.~Weinberg, {\em {The quantum theory of fields. Vol. 2: Modern applications}}.
\newblock Cambridge University Press, 8, 2013.

\bibitem{Volkov:1973vd}
D.~V. Volkov, ``{Phenomenological Lagrangians},'' {\em Fiz. Elem. Chast. Atom.
  Yadra} {\bfseries 4} (1973) 3--41.

\bibitem{Ivanov:1975zq}
E.~Ivanov and V.~Ogievetsky, ``{The Inverse Higgs Phenomenon in Nonlinear
  Realizations},'' \href{http://dx.doi.org/10.1007/BF01028947}{{\em Teor. Mat.
  Fiz.} {\bfseries 25} (1975) 164--177}.

\bibitem{Salam:1970qk}
A.~Salam and J.~Strathdee, ``{Nonlinear realizations. 2. Conformal symmetry},''
  \href{http://dx.doi.org/10.1103/PhysRev.184.1760}{{\em Phys. Rev.} {\bfseries
  184} (1969) 1760--1768}.

\bibitem{Isham:1970xz}
C.~Isham, A.~Salam, and J.~Strathdee, ``{Broken chiral and conformal symmetry
  in an effective-lagrangian formalism},''
  \href{http://dx.doi.org/10.1103/PhysRevD.2.685}{{\em Phys. Rev. D} {\bfseries
  2} (1970) 685--690}.

\bibitem{Isham:1970gz}
C.~Isham, A.~Salam, and J.~Strathdee, ``{Spontaneous breakdown of conformal
  symmetry},'' \href{http://dx.doi.org/10.1016/0370-2693(70)90177-2}{{\em Phys.
  Lett. B} {\bfseries 31} (1970) 300--302}.

\bibitem{Low:2001bw}
I.~Low and A.~V. Manohar, ``{Spontaneously broken space-time symmetries and
  Goldstone's theorem},''
  \href{http://dx.doi.org/10.1103/PhysRevLett.88.101602}{{\em Phys. Rev. Lett.}
  {\bfseries 88} (2002) 101602},
  \href{http://arxiv.org/abs/hep-th/0110285}{{\ttfamily arXiv:hep-th/0110285}}.

\bibitem{McArthur:2010zm}
I.~McArthur, ``{Nonlinear realizations of symmetries and unphysical Goldstone
  bosons},'' \href{http://dx.doi.org/10.1007/JHEP11(2010)140}{{\em JHEP}
  {\bfseries 11} (2010) 140}, \href{http://arxiv.org/abs/1009.3696}{{\ttfamily
  arXiv:1009.3696 [hep-th]}}.

\bibitem{Hinterbichler:2012mv}
K.~Hinterbichler, A.~Joyce, and J.~Khoury, ``{Non-linear Realizations of
  Conformal Symmetry and Effective Field Theory for the Pseudo-Conformal
  Universe},'' \href{http://dx.doi.org/10.1088/1475-7516/2012/06/043}{{\em
  JCAP} {\bfseries 06} (2012) 043},
  \href{http://arxiv.org/abs/1202.6056}{{\ttfamily arXiv:1202.6056 [hep-th]}}.

\bibitem{DHoker:1994rdl}
E.~D'Hoker and S.~Weinberg, ``{General effective actions},''
  \href{http://dx.doi.org/10.1103/PhysRevD.50.R6050}{{\em Phys. Rev. D}
  {\bfseries 50} (1994) 6050--6053},
  \href{http://arxiv.org/abs/hep-ph/9409402}{{\ttfamily arXiv:hep-ph/9409402}}.

\bibitem{Komargodski:2012ek}
Z.~Komargodski and A.~Zhiboedov, ``{Convexity and Liberation at Large Spin},''
  \href{http://dx.doi.org/10.1007/JHEP11(2013)140}{{\em JHEP} {\bfseries 1311}
  (2013) 140},
\href{http://arxiv.org/abs/1212.4103}{{\ttfamily arXiv:1212.4103 [hep-th]}}.
%%CITATION = ARXIV:1212.4103;%%.

\bibitem{Fitzpatrick:2012yx}
A.~L. Fitzpatrick, J.~Kaplan, D.~Poland, and D.~Simmons-Duffin, ``{The Analytic
  Bootstrap and AdS Superhorizon Locality},''
  \href{http://dx.doi.org/10.1007/JHEP12(2013)004}{{\em JHEP} {\bfseries 1312}
  (2013) 004},
\href{http://arxiv.org/abs/1212.3616}{{\ttfamily arXiv:1212.3616 [hep-th]}}.
%%CITATION = ARXIV:1212.3616;%%.

\bibitem{Costa:2017twz}
M.~S. Costa, T.~Hansen, and J.~Penedones, ``{Bounds for OPE coefficients on the
  Regge trajectory},'' \href{http://dx.doi.org/10.1007/JHEP10(2017)197}{{\em
  JHEP} {\bfseries 10} (2017) 197},
  \href{http://arxiv.org/abs/1707.07689}{{\ttfamily arXiv:1707.07689
  [hep-th]}}.

\bibitem{Afkhami-Jeddi:2017rmx}
N.~Afkhami-Jeddi, T.~Hartman, S.~Kundu, and A.~Tajdini, ``{Shockwaves from the
  Operator Product Expansion},''
  \href{http://dx.doi.org/10.1007/JHEP03(2019)201}{{\em JHEP} {\bfseries 03}
  (2019) 201},
\href{http://arxiv.org/abs/1709.03597}{{\ttfamily arXiv:1709.03597 [hep-th]}}.
%%CITATION = ARXIV:1709.03597;%%.

\bibitem{Afkhami-Jeddi:2018own}
N.~Afkhami-Jeddi, S.~Kundu, and A.~Tajdini, ``{A Conformal Collider for
  Holographic CFTs},'' \href{http://dx.doi.org/10.1007/JHEP10(2018)156}{{\em
  JHEP} {\bfseries 10} (2018) 156},
\href{http://arxiv.org/abs/1805.07393}{{\ttfamily arXiv:1805.07393 [hep-th]}}.
%%CITATION = ARXIV:1805.07393;%%.

\bibitem{Kulaxizi:2017ixa}
M.~Kulaxizi, A.~Parnachev, and A.~Zhiboedov, ``{Bulk Phase Shift, CFT Regge
  Limit and Einstein Gravity},''
  \href{http://dx.doi.org/10.1007/JHEP06(2018)121}{{\em JHEP} {\bfseries 06}
  (2018) 121}, \href{http://arxiv.org/abs/1705.02934}{{\ttfamily
  arXiv:1705.02934 [hep-th]}}.

\bibitem{Intriligator:2003jj}
K.~A. Intriligator and B.~Wecht, ``{The Exact superconformal R symmetry
  maximizes a},'' \href{http://dx.doi.org/10.1016/S0550-3213(03)00459-0}{{\em
  Nucl. Phys. B} {\bfseries 667} (2003) 183--200},
  \href{http://arxiv.org/abs/hep-th/0304128}{{\ttfamily arXiv:hep-th/0304128}}.

\bibitem{Kutasov:2003iy}
D.~Kutasov, A.~Parnachev, and D.~A. Sahakyan, ``{Central charges and U(1)(R)
  symmetries in N=1 superYang-Mills},''
  \href{http://dx.doi.org/10.1088/1126-6708/2003/11/013}{{\em JHEP} {\bfseries
  11} (2003) 013}, \href{http://arxiv.org/abs/hep-th/0308071}{{\ttfamily
  arXiv:hep-th/0308071}}.

\bibitem{Kutasov:2003ux}
D.~Kutasov, ``{New results on the 'a theorem' in four-dimensional
  supersymmetric field theory},''
  \href{http://arxiv.org/abs/hep-th/0312098}{{\ttfamily arXiv:hep-th/0312098}}.

\bibitem{Pham:1985cr}
T.~Pham and T.~N. Truong, ``{Evaluation of the Derivative Quartic Terms of the
  Meson Chiral Lagrangian From Forward Dispersion Relation},''
  \href{http://dx.doi.org/10.1103/PhysRevD.31.3027}{{\em Phys. Rev. D}
  {\bfseries 31} (1985) 3027}.

\bibitem{Qi1}
L.~Qi, H.-H. Dai, and D.~Han, ``Conditions for strong ellipticity and
  m-eigenvalues,'' \href{http://dx.doi.org/10.1007/s11464-009-0016-6}{{\em
  Frontiers of Mathematics in China} {\bfseries 4} no.~2, (2009) 349}.
  \url{https://doi.org/10.1007/s11464-009-0016-6}.

\bibitem{Qi2}
C.~Ling, J.~Nie, L.~Qi, and Y.~Ye, ``Biquadratic optimization over unit spheres
  and semidefinite programming relaxations,''
  \href{http://dx.doi.org/10.1137/080729104}{{\em SIAM Journal on Optimization}
  {\bfseries 20} no.~3, (2010) 1286--1310},
  \href{http://arxiv.org/abs/https://doi.org/10.1137/080729104}{{\ttfamily
  https://doi.org/10.1137/080729104}}. \url{https://doi.org/10.1137/080729104}.

\bibitem{Elvang:2012yc}
H.~Elvang and T.~M. Olson, ``{RG flows in d dimensions, the dilaton effective
  action, and the a-theorem},''
  \href{http://dx.doi.org/10.1007/JHEP03(2013)034}{{\em JHEP} {\bfseries 03}
  (2013) 034}, \href{http://arxiv.org/abs/1209.3424}{{\ttfamily arXiv:1209.3424
  [hep-th]}}.

\end{thebibliography}\endgroup

\end{document}